\documentclass[10pt,letterpaper]{article}

\usepackage{atbegshi}
\AtBeginDocument{\AtBeginShipoutNext{\AtBeginShipoutDiscard}}

\usepackage{graphicx}
\graphicspath{ {./images/} }

\usepackage{wrapfig}
\usepackage{blindtext}

\usepackage[switch]{lineno} 

\usepackage{cogsci}
\usepackage{pslatex}
\usepackage{apacite}  
\AtBeginDocument{}

\usepackage[OT1]{fontenc}
\usepackage[font=small,labelfont=bf]{caption}

\usepackage{lmodern}

\fontfamily{cmr}

\usepackage{physics}
\usepackage{amsmath,amssymb}

\usepackage{caption}
\usepackage{subcaption}
\usepackage{graphicx}

\usepackage[most]{tcolorbox}

\usepackage{dblfloatfix}

\DeclareGraphicsExtensions{.png,.pdf}

\newcommand\FIGURESCALE{0.28}  
\newcommand\SUBFIGUREWIDTH{0.35}  

\title{Successes and critical failures of neural networks in capturing human-like speech recognition}
 
\author{{\large \bf Federico Adolfi} \\
  Ernst Strüngmann Institute (ESI) for Neuroscience in Cooperation with Max Planck Society, Frankfurt, Germany \\
  University of Bristol, School of Psychological Science, Bristol, United Kingdom \\
   \AND {\large \bf Jeffrey S. Bowers} \\
  University of Bristol, School of Psychological Science, Bristol, United Kingdom \\
  \AND {\large \bf David Poeppel} \\
  Ernst Strüngmann Institute (ESI) for Neuroscience in Cooperation with Max Planck Society, Frankfurt, Germany \\
  Department of Psychology, New York University, New York, United States \\
  Max Planck NYU Center for Language, Music, and Emotion, Frankfurt, Germany, New York, NY
  }

\begin{document}

\makeatletter
\newcommand{\settitle}{\@maketitle}. 
\makeatother

\begin{titlepage}
\settitle
\paragraph{
Correspondence should be addressed to Federico Adolfi (fedeadolfi@bristol.ac.uk)
}
\end{titlepage}

\maketitle
\begin{abstract}
Natural and artificial audition can in principle acquire different solutions to a given problem. The constraints of the task, however, can nudge the cognitive science and engineering of audition to qualitatively converge, suggesting that a closer mutual examination would potentially enrich artificial hearing systems and process models of the mind and brain. Speech  recognition — an area ripe for such exploration — is inherently robust in humans to a number transformations at various spectrotemporal granularities. To what extent are these robustness  profiles  accounted  for  by  high-performing neural network systems? We bring together experiments in speech recognition under a single synthesis framework to  evaluate  state-of-the-art  neural networks  as  stimulus-computable,  optimized  observers. In a series of experiments, we (1) clarify how influential speech manipulations in the literature relate to each other and to natural speech, (2) show the granularities at which machines exhibit out-of-distribution robustness, reproducing classical perceptual phenomena in humans, (3) identify the specific conditions where model predictions of human performance differ, and (4) demonstrate a crucial failure of all artificial systems to perceptually recover where humans do, suggesting alternative directions for theory and model building. These findings encourage a tighter synergy between the cognitive science and engineering of audition.

\end{abstract}


\section{Introduction}

\begin{figure}[!ht]
    \begin{center}
    \includegraphics[scale=0.72]{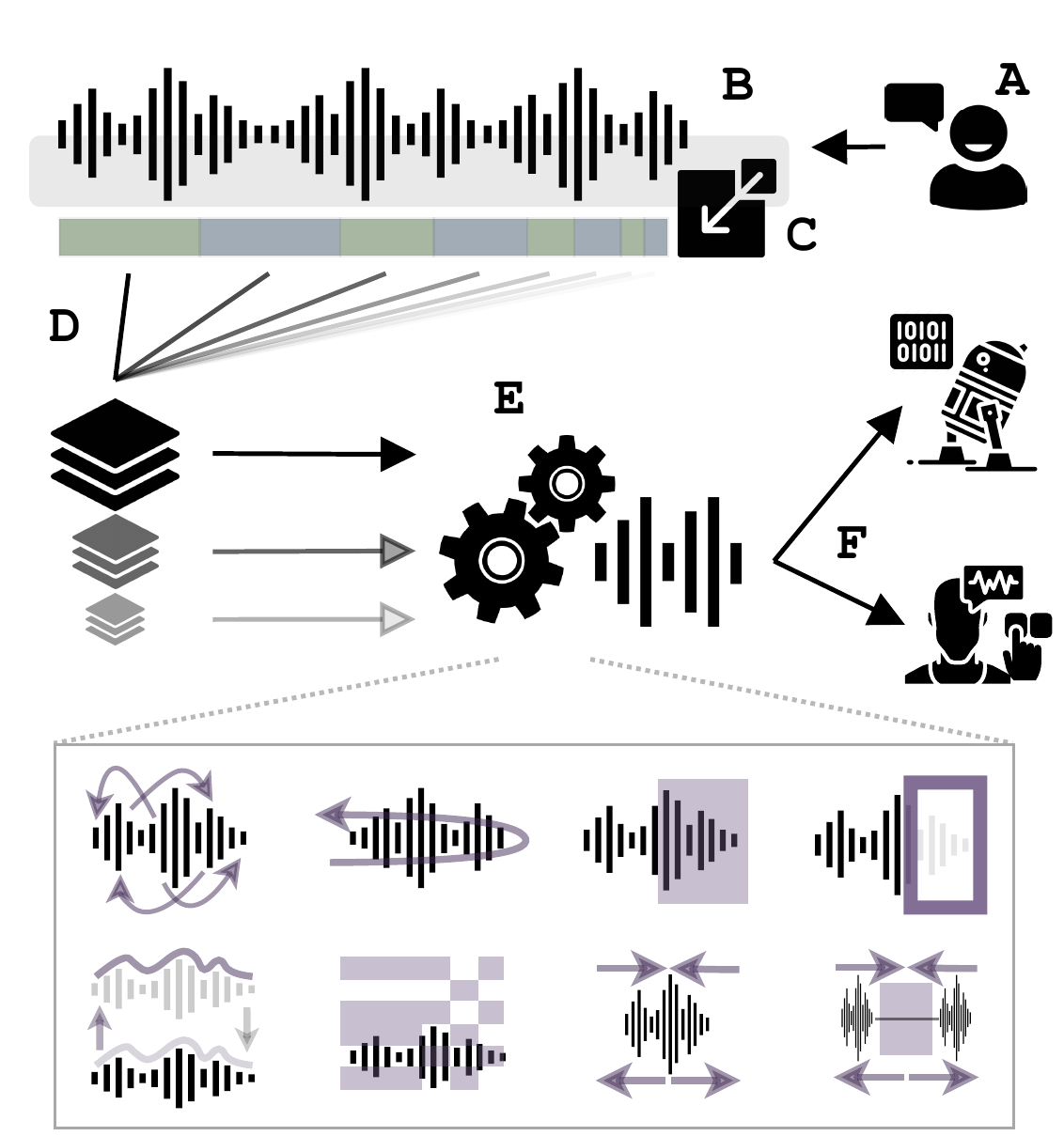}
    \end{center}
    \caption{Human speech (A) is recorded and represented as a 1-dimensional signal in the time domain (B), which is optionally converted to a spectrogram-like representation in the time-frequency domain (C). It is subsequently segmented in parallel at various spectrotemporal scales (D). The resulting slices become the input to a transformation (E) — which may involve shuffling, reversing, masking, silencing, chimerizing, mosaicizing, time warping, or repackaging. Finally, the outputs are sequenced and the resulting time-domain signals are presented to both humans and optimized observer models (F).}
    \label{figure:experiments}
\end{figure}

Audition systems — artificial and biological — can in principle acquire qualitatively different solutions to the same ecological problem. 
For instance, redundancy at the input or lack thereof, relative to the structure and complexity of the problem, can encourage systems towards divergent or convergent evolution. 
Whether performance-optimized engineering solutions and biological perception converge for a particular problem determines, in part, the extent to which artificial auditory systems can play a role as process models of the mind and brain \cite{ma2020NeuralNetworkWalksacq}. 

Although neural networks for audio have achieved remarkable performance in tasks such as speech recognition, most of the links to computational cognitive science have come from vision, with audition being comparatively neglected \cite{cichy2019DeepNeuralNetworkstics}.
Audition as a field has its own set of unique challenges: explaining and building systems that must integrate sound information at various spectrotemporal scales to accomplish even the most basic recognition task \cite{poeppel2008SpeechPerceptionInterfaceptrsb, poeppel2020SpeechRhythmsTheirnrna}.
Nevertheless, research into audition can avoid pitfalls in model evaluation by looking at emerging critiques of neural networks for vision \cite{bowersDeepProblemsNeural2022} and adopting a more qualitative and diverse approach \cite{navarroDevilDeepBlue2019}. We therefore set out to characterize the solutions acquired by machine hearing systems as compared to humans, drawing bridges across influential research lines in auditory cognitive science and engineering.

An area of audition where the two disciplines once worked in close allegiance is speech recognition. 
The engineering of machine hearing has produced a zoo of task-optimized architectures — convolutional \cite{veysov2020towardimagenetstt}, recurrent \cite{amodeiDeepSpeechEndtoEnd,hannun2014DeepSpeechScalingac}, and more recently, transformer-based \cite{baevskiWav2vecFrameworkSelfSupervised,schneider2019Wav2vecUnsupervisedPretrainingac} — achieving performance levels impressive enough (on benchmark tasks) to afford numerous real-world applications. The cognitive science of audition provides a complementary perspective from biological hearing. 
A research program based on multi-scale perturbations to natural signals — going back to the 1950s \cite{miller1950IntelligibilityInterruptedSpeech}, active through decades \cite{saberi1999CognitiveRestorationReversedn, smith2002ChimaericSoundsRevealn,shannon1995SpeechRecognitionPrimarilysa}, and still thriving \cite{gotoh2017EffectPermutationsTimetjotasoa,ueda2017IntelligibilityLocallyTimereversedsr}, has provided detailed descriptions of performance patterns in humans. The question is whether these engineering and scientific insights converge, and to what extent they can more explicitly inform each other.

Speech recognition in humans is inherently resistant to a number of perturbations at various granularities, exhibiting a form of out-of-distribution robustness analogous to how biological (but typically not artificial) vision generalizes to contour images and other transformations \cite{evans2021BiologicalConvolutionsImproveb}. 
This has been uncovered by a large set of experiments which process natural speech in a selective manner at multiple spectrotemporal scales (e.g., \citeNP{saberi1999CognitiveRestorationReversedn,smith2002ChimaericSoundsRevealn,shannon1995SpeechRecognitionPrimarilysa}). 
The results are suggestive of the properties of mid-level stages of audition that drive any downstream task such as prediction and categorization. 
Are these robustness profiles accounted for by modern neural network systems?

We make explicit the synthesis space implied by these experiments, bringing them together under a single framework (Fig.~\ref{figure:experiments}) that allows us to simulate behavior exhaustively in search for human-machine alignment.
By this we mean that each classical experiment implicitly defines a space of possible simulations given by the experimental parameters (e.g., the temporal scale at which perturbations are performed).
We combine and vary these in order to cover more ground than what was the case in the original experiments.
In this way we can give the qualitatively human-like performance patterns a chance to emerge in the results without limiting their manifestation to the narrow parameter range of past studies.

The broader rationale is that insights about a perceptual system and its input signal (in this case, speech) can be gleaned by observing the transformations and spectrotemporal granularities for which systems show perturbation-robust behavior.
Systems will show performance curves reflecting whether (a) they rely on perturbation-invariant transformations at various granularities, and (b) information evolving at these scales is present and relevant for the downstream task. These in turn depend on the relevant signal cues being unique such that all solutions, artificial or biological, tend towards exploiting it.
With this framework in place, we perform multi-scale audio analysis and synthesis, evaluate state-of-the-art neural networks as stimulus-computable optimized observers, and compare the simulated predictions to human performance.

This paper is organized as follows.
First, we clarify how the different audio manipulations in the literature relate to each other by describing their effects in a common space: the sparsity statistics of the input.
This allows us to link the distribution of experimental stimuli in human cognitive science to that of training and testing examples for artificial systems. Synthetic and natural speech fill this space and show regions where human and machine performance is robust outside the training distribution. Second, in a series of experiments we find that, while several classical perceptual phenomena are well-predicted by high-performing, speech-to-text neural network architectures, more destructive perturbations reveal differences amongst models and between these and humans. Finally, we demonstrate a systematic failure of all artificial systems to perceptually recover where humans do, which is suggestive of alternative directions for theorizing, computational cognitive modeling, and, more speculatively, improvement of engineering solutions. 

\section{Results}

\begin{figure*}[!hb]
    \begin{center}
    \includegraphics[scale=1.0]{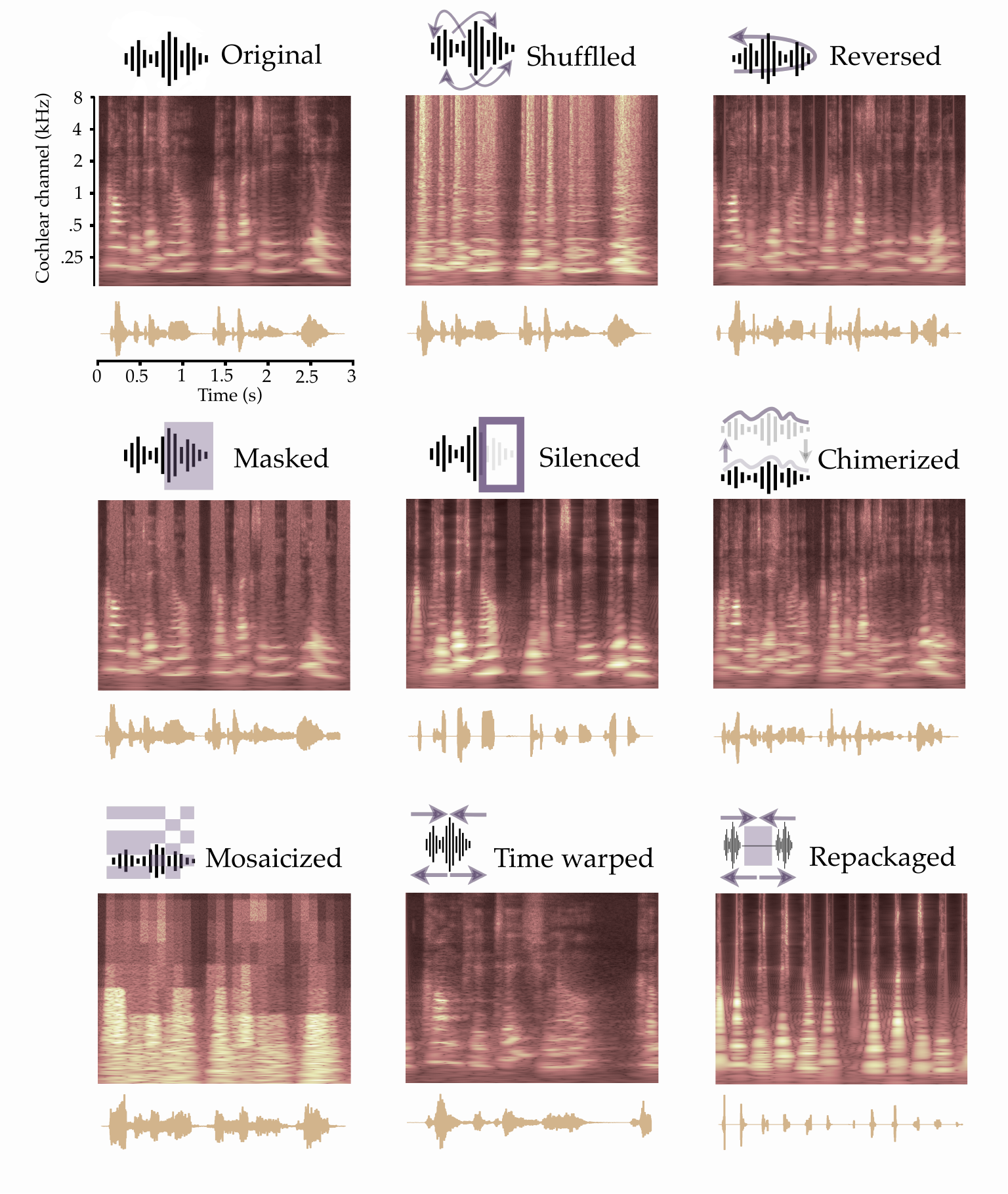}
    \end{center}
    \caption{
    Spectrogram and waveform representations of natural and resynthesized speech for all perturbations of a single 3-second utterance: ``computational complexity theory". To illustrate the effect of various perturbations on the signal, we show moderate perturbation magnitudes: \textit{shuffling} is done at a 2 ms timescale; \textit{reversing} at 150 ms; \textit{masking} and \textit{silencing} are done at 300 ms; \textit{chimerizing} is done with 30 bands and targeting the envelopes for reversal at 100 ms; \textit{mosaicizing} is done with 60 bands and a frequency window of 6 bands and time window of 100 ms; \textit{time warping} is applied with a warp factor of 0.5 (stretching); \textit{repackaging} is done with a warp factor of 3.0 (compressing), a time window of 250 ms and an insertion of silence of 167ms. Refer to the main text and Methods section for details on the audio perturbations and resynthesis procedures.
    }
    \label{figure:perturbations-demo}
\end{figure*}

We characterize the input space and report performance on speech recognition, measured by the word error rate (WER), for multiple experiments with trained neural networks including convolutional, recurrent and transformer models (see Methods for details).
Our experimental framework (Fig. \ref{figure:experiments}) systematizes and integrates classical speech perturbations.
These re-synthesis procedures split the signal into segments and apply a transformation within each segment, such as \textit{shuffling}, \textit{reversing}, \textit{masking}, \textit{silencing}, \textit{chimerizing}, \textit{mosaicizing}, \textit{time warping}, or \textit{repackaging} (see Fig. \ref{figure:perturbations-demo} for example spectrograms of natural and perturbed signals, and Methods section for details).
Then the segments are concatenated together and the resulting perturbed speech is presented to machines. The performance of the models under different perturbations is therefore evaluated and plotted separately at various scales and perturbation parameter values.

The rationale for choosing these perturbations, which are not variants of natural speech, is that (i) they represent a cohesive family of manipulations to the speech signal with well-known human performance profiles; (ii) they represent a unique opportunity to test for out-of-distribution
robustness/generalization, as humans are robust to these perturbations at specific timescales without having been trained explicitly; and (iii) they each allow informative interpretations of the results in terms of (a) the specific invariances learned by neural networks and (b) the timescales at which these invariances operate. For instance, if a trained model's performance is unaffected by a specific perturbation at time scale X (e.g., 250 ms) which destroys the structure of feature Y (e.g., phase spectrum) but preserves that of feature Z (e.g., magnitude spectrum), then we can infer that the transformation learned by the model is likely invariant in this particular sense.

To avoid pervasive problems \cite{bowersDeepProblemsNeural2022,dujmovicPitfallsMeasuringRepresentational2022a,guestLogicalInferenceBrains2023} with monolithic, quantitative assessments of predictive accuracy (e.g., a single brain activity prediction score), in this work we focus instead on the qualitative fit \cite{navarroDevilDeepBlue2019,bowersDeepProblemsNeural2022} between machines and humans.

That is, we first identify the canonical performance curve exhibited by humans in response to parametric speech perturbations, and then we search for this pattern by systematic visual inspection in the performance profile of neural networks across many combinations of experimental parameters, including the original one used in human studies.
For instance, if humans exhibit a U-shaped performance curve as a perturbation parameter value is increased, we search for such a curve in the performance profile of neural network models.
In all cases, we plot the results on axes chosen to match the classical experiments we build on, to facilitate comparisons.

The main results summarizing the findings of our more comprehensive evaluation are presented here succinctly and later discussed more comprehensively.

\subsection{Input statistics: sparsity and out-of-distribution robustness}

\begin{figure}[!ht]
    \begin{center}
    \includegraphics[scale=0.28]{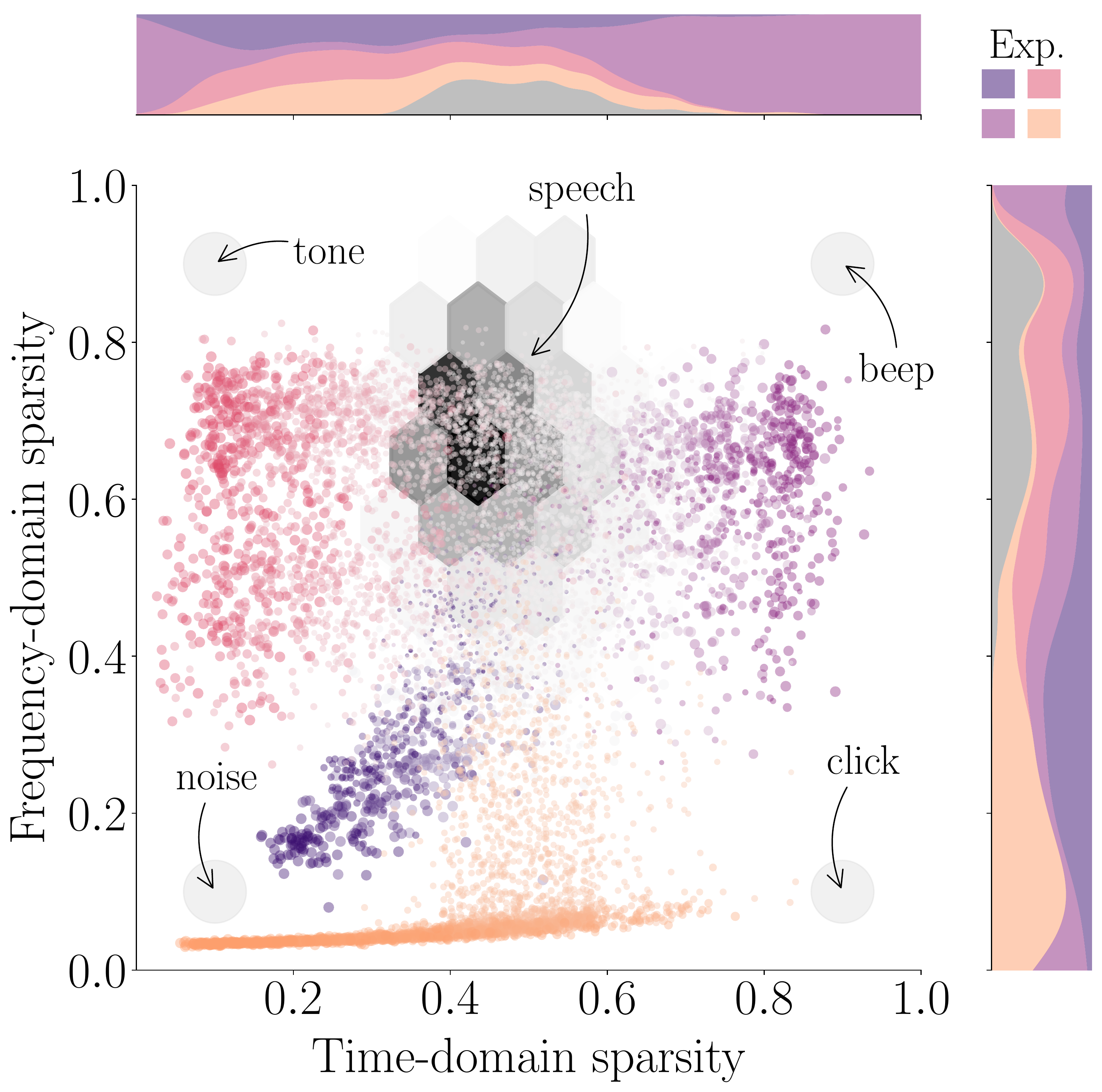}
    \end{center}
    \caption{Schematic of how natural and experimental distributions fill the input space defined by sparsity in time and frequency. The natural speech distribution is shown in grayscale hexagons located at the center. A subset of the processed audio samples are shown in color according to 4 example experiments (color code on the top right). Each dot represents a speech utterance that has been perturbed according to an example resynthesis procedure (here \textit{shuffling [orange], masking [purple], silencing [violet]} and \textit{mosaicizing [red]}; see main text and Methods for details). The perturbed signal is run through the sparsity analysis, obtaining one value for time sparsity and another for frequency sparsity. Hue and size indicates the magnitude of the perturbation according to its respective parameter set (e.g., window length). Marginal distributions are 'filled' such that the proportion of samples for different experiments is reflected at each point. It can be seen that audio transformations systematically push samples away from the training set. Canonical signals (noise, tone, beep, click) are annotated at the extremes for reference. The sparsity plots for each perturbation are reported later individually.}
    \label{figure:gini}
\end{figure}

\begin{figure*}[hb!]
    \begin{center}
        \centering
        
        \begin{subfigure}[t]{\SUBFIGUREWIDTH\textwidth}
            \centering
            \subcaption{}
            \includegraphics[scale=\FIGURESCALE]{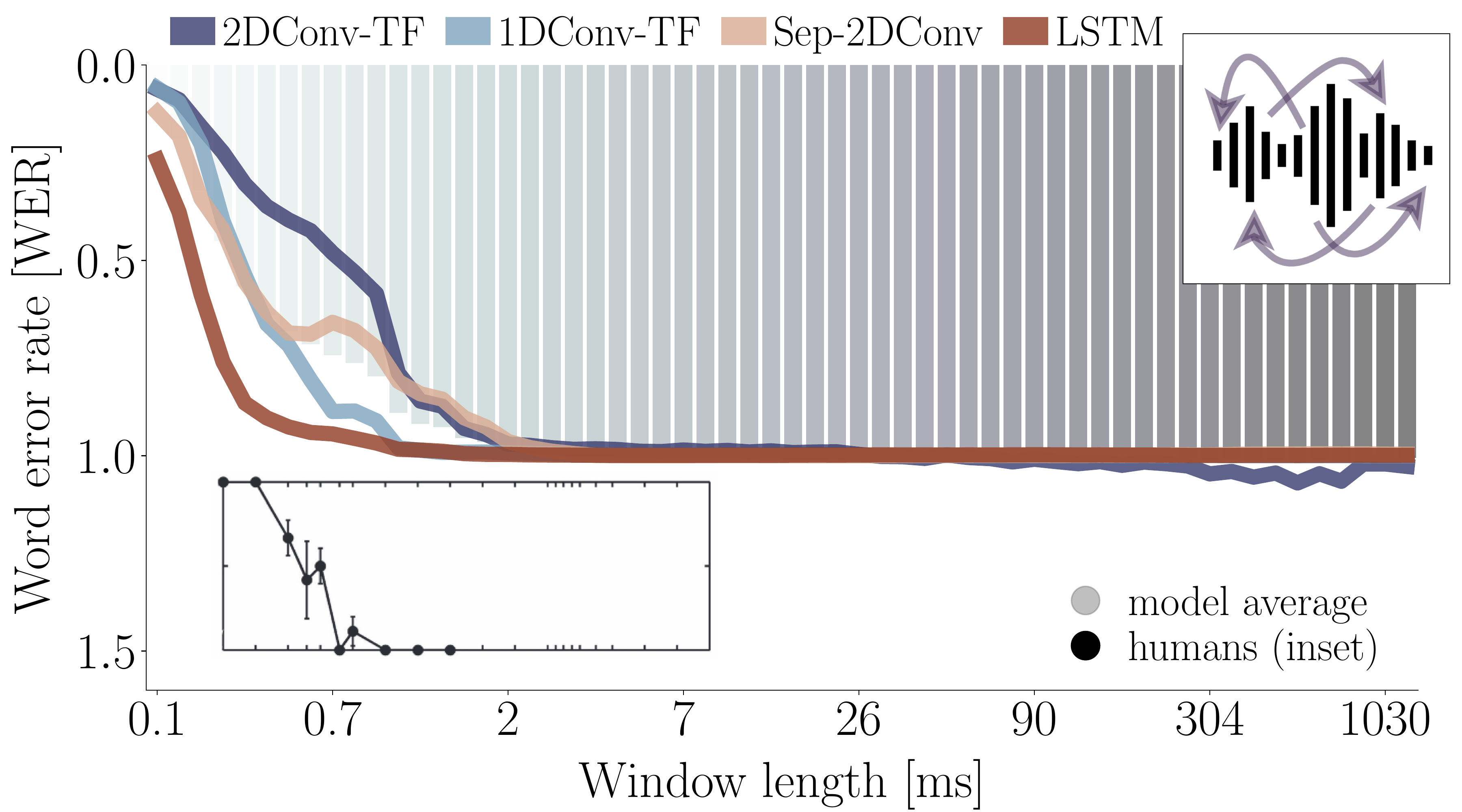}
        \end{subfigure}
        \hfill
        \begin{subfigure}[t]{\SUBFIGUREWIDTH\textwidth}
            \centering
            \subcaption{}
            \includegraphics[scale=\FIGURESCALE]{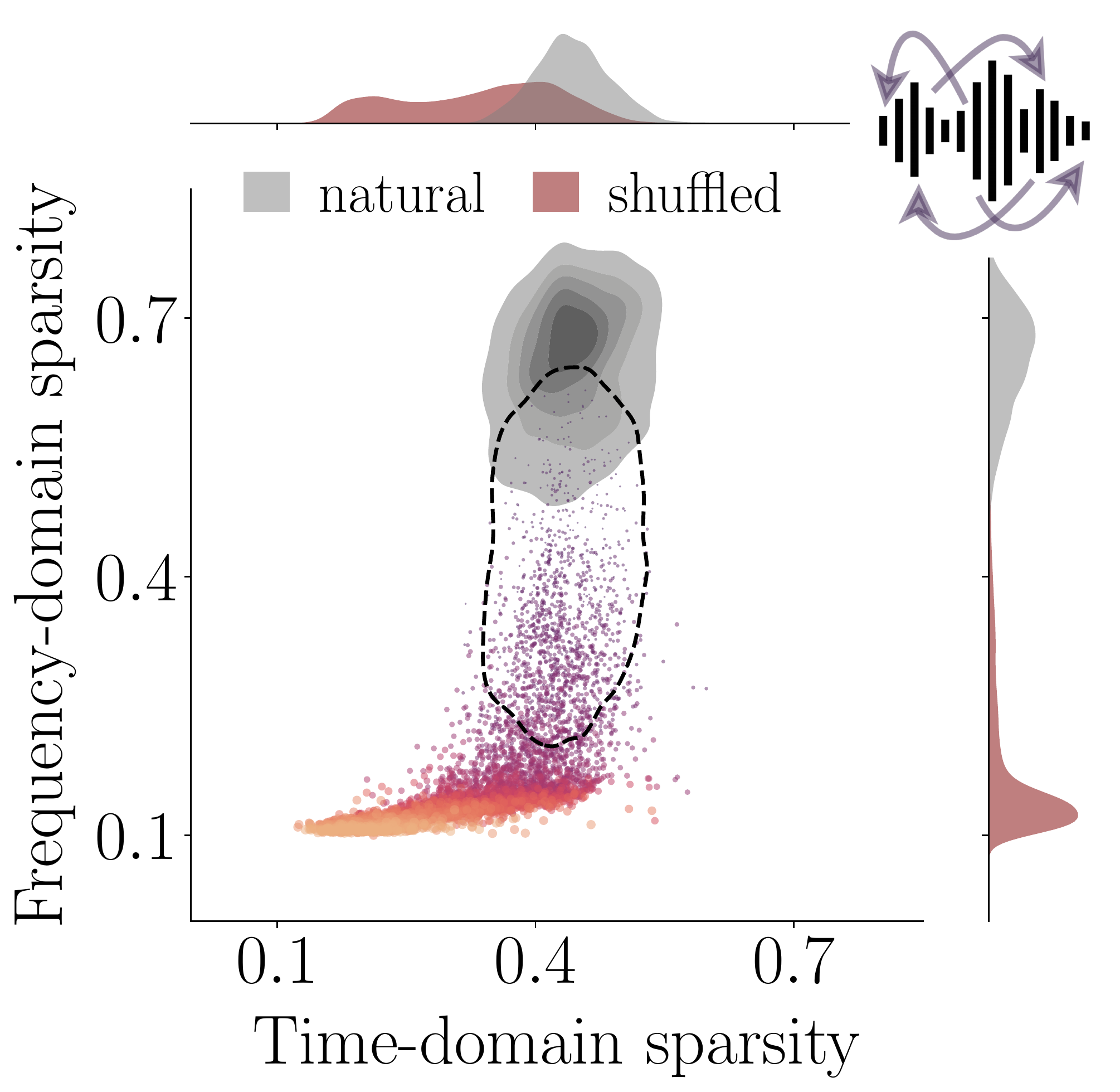}
        \end{subfigure}
    
        \vfill
        
        \begin{subfigure}[t]{\SUBFIGUREWIDTH\textwidth}
            \centering
            \subcaption{}
            \includegraphics[scale=\FIGURESCALE]{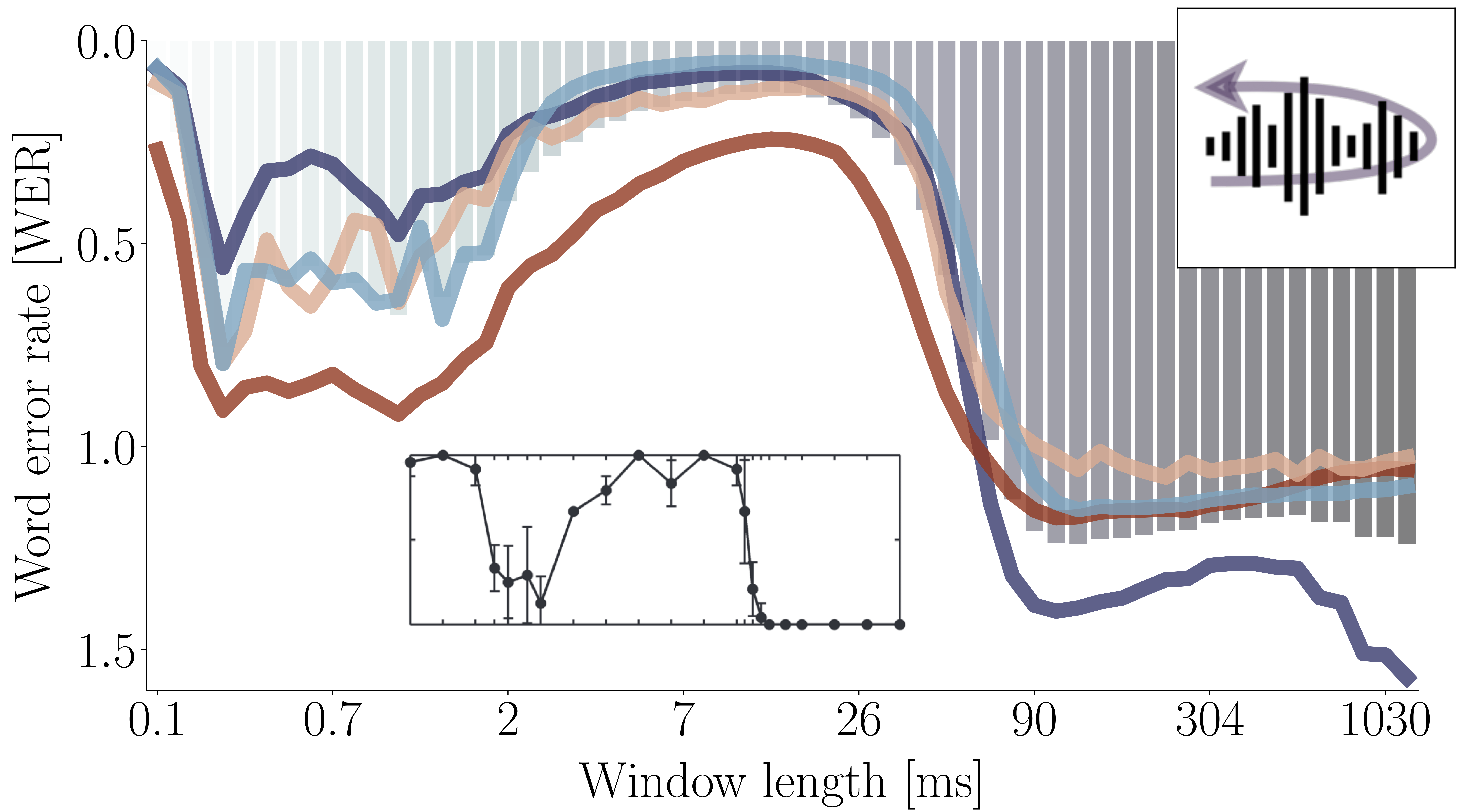}
        \end{subfigure}
        \hfill
        \begin{subfigure}[t]{\SUBFIGUREWIDTH\textwidth}
            \centering
            \subcaption{}
            \includegraphics[scale=\FIGURESCALE]{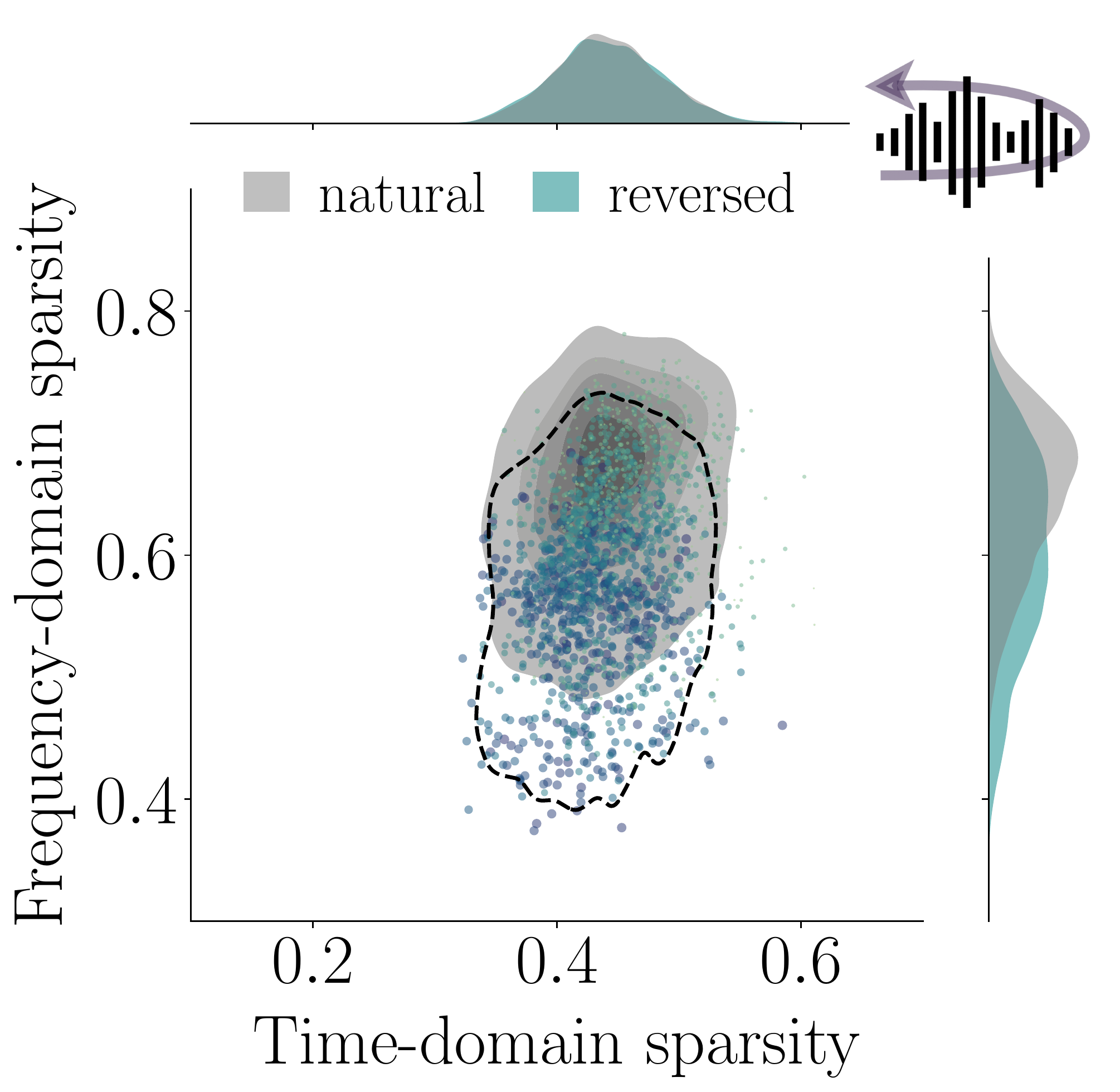}
        \end{subfigure}
        
        \vfill
        
        \begin{subfigure}[t]{\SUBFIGUREWIDTH\textwidth}
            \centering
            \subcaption{}
            \includegraphics[scale=\FIGURESCALE]{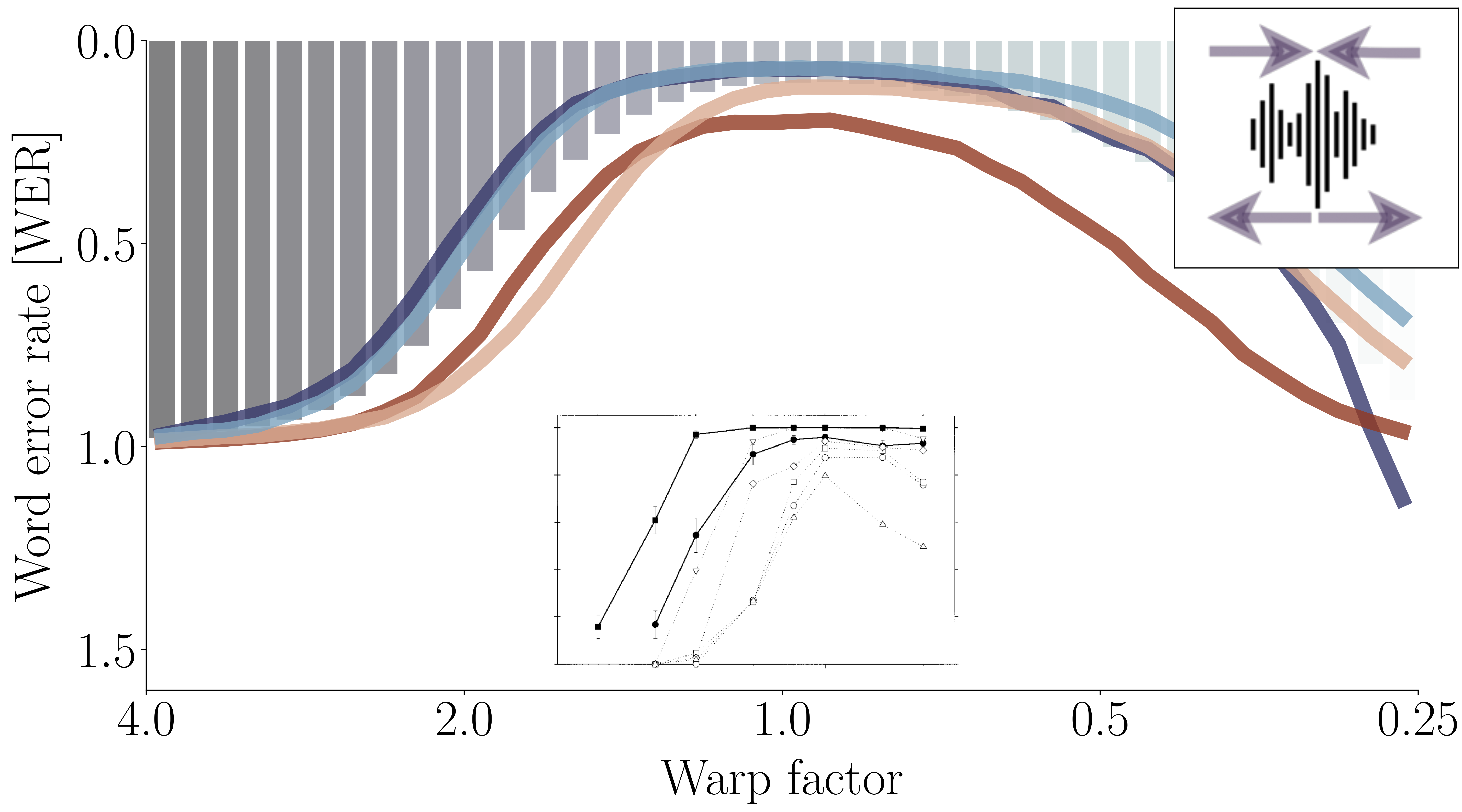}
        \end{subfigure}
        \hfill
        \begin{subfigure}[t]{\SUBFIGUREWIDTH\textwidth}
            \centering
            \subcaption{}
            \includegraphics[scale=\FIGURESCALE]{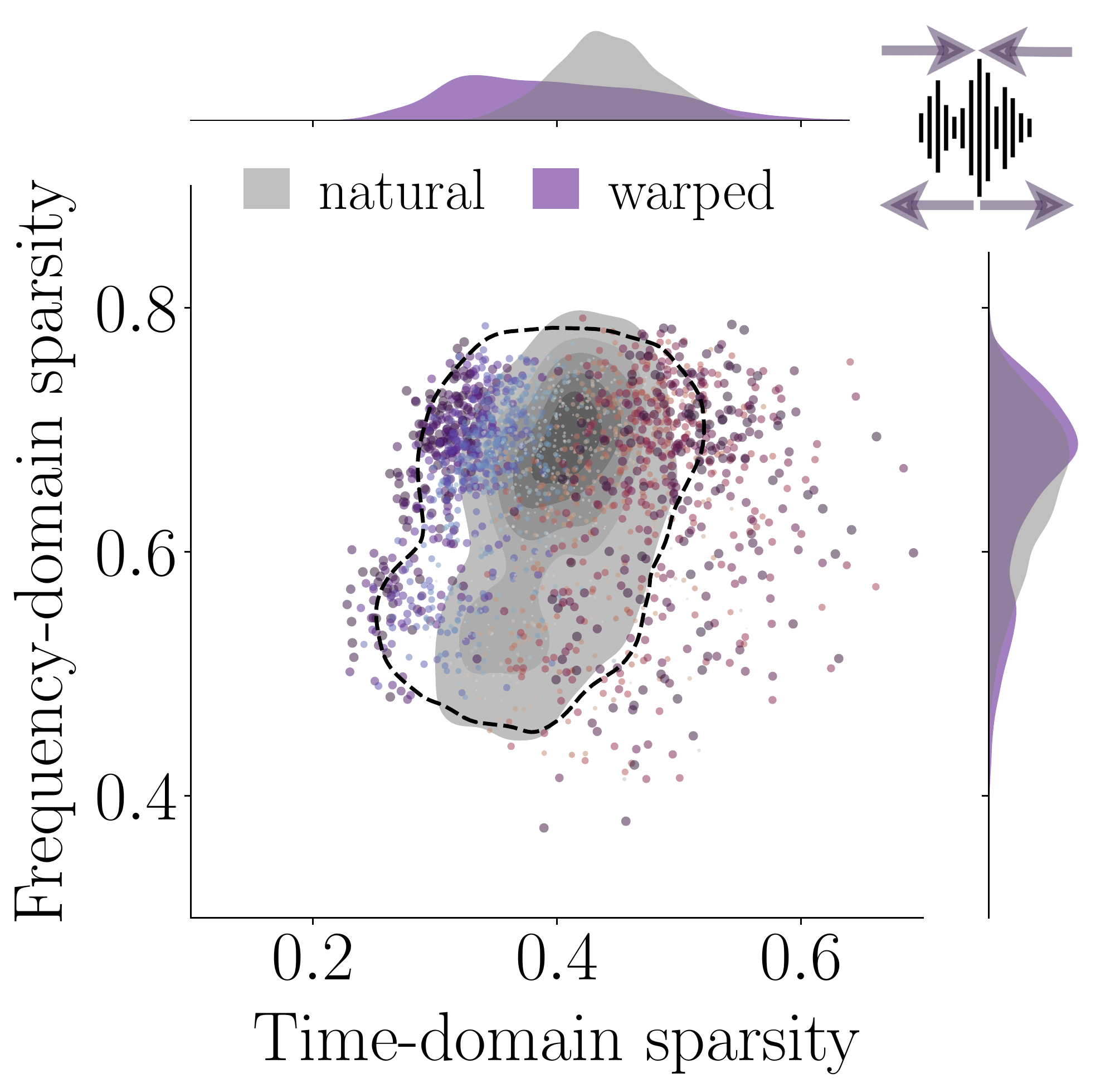}
        \end{subfigure}

    \end{center}
    
\caption{Machines and humans are resistant to (A) shuffling, (C) reversal, and (E) time warping, at comparable granularities, exhibiting qualitatively similar patterns of out-of-distribution robustness (insets adapted from \citeNP{gotoh2017EffectPermutationsTimetjotasoa, fu2001RecognitionTimedistortedSentencesjasa} have x/y-axis ranges comparable to the corresponding main graphs; inset on panel E shows performance for normal hearing listeners [filled shapes] and cochlear implant users [blank shapes]). Color coding of models is indicated in panel A. We plot performance (WER) as a function of perturbation timescale (window length in ms) for shuffling and reversal, and as a function of warp factor for time warping (left-hand panels). The effect of the manipulations on the input distributions (right-hand panels) is visualized with hues and sizes representing synthesis parameters (B: window length, D: window length, F: warp factor, respectively; dashed contour shows region of 85-100\% model performance).}
\label{figure:convergence}
\end{figure*}

\begin{figure*}[hb!]
    \begin{center}
        \centering
        
        \begin{subfigure}[t]{\SUBFIGUREWIDTH\textwidth}
            \centering
            \subcaption{}
            \includegraphics[scale=\FIGURESCALE]{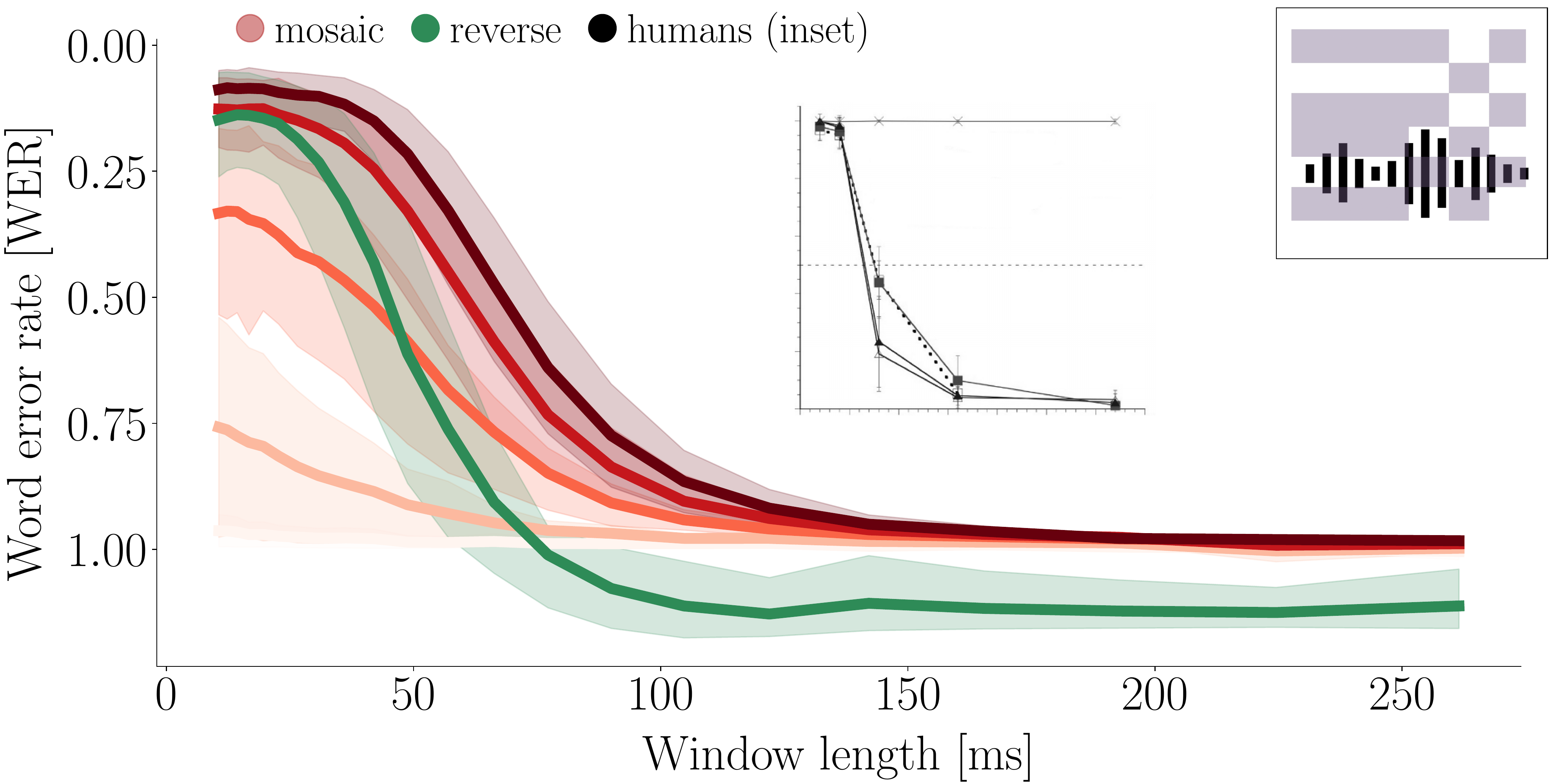}
        \end{subfigure}
            \hfill
        \begin{subfigure}[t]{\SUBFIGUREWIDTH\textwidth}
            \centering
            \subcaption{}
            \includegraphics[scale=\FIGURESCALE]{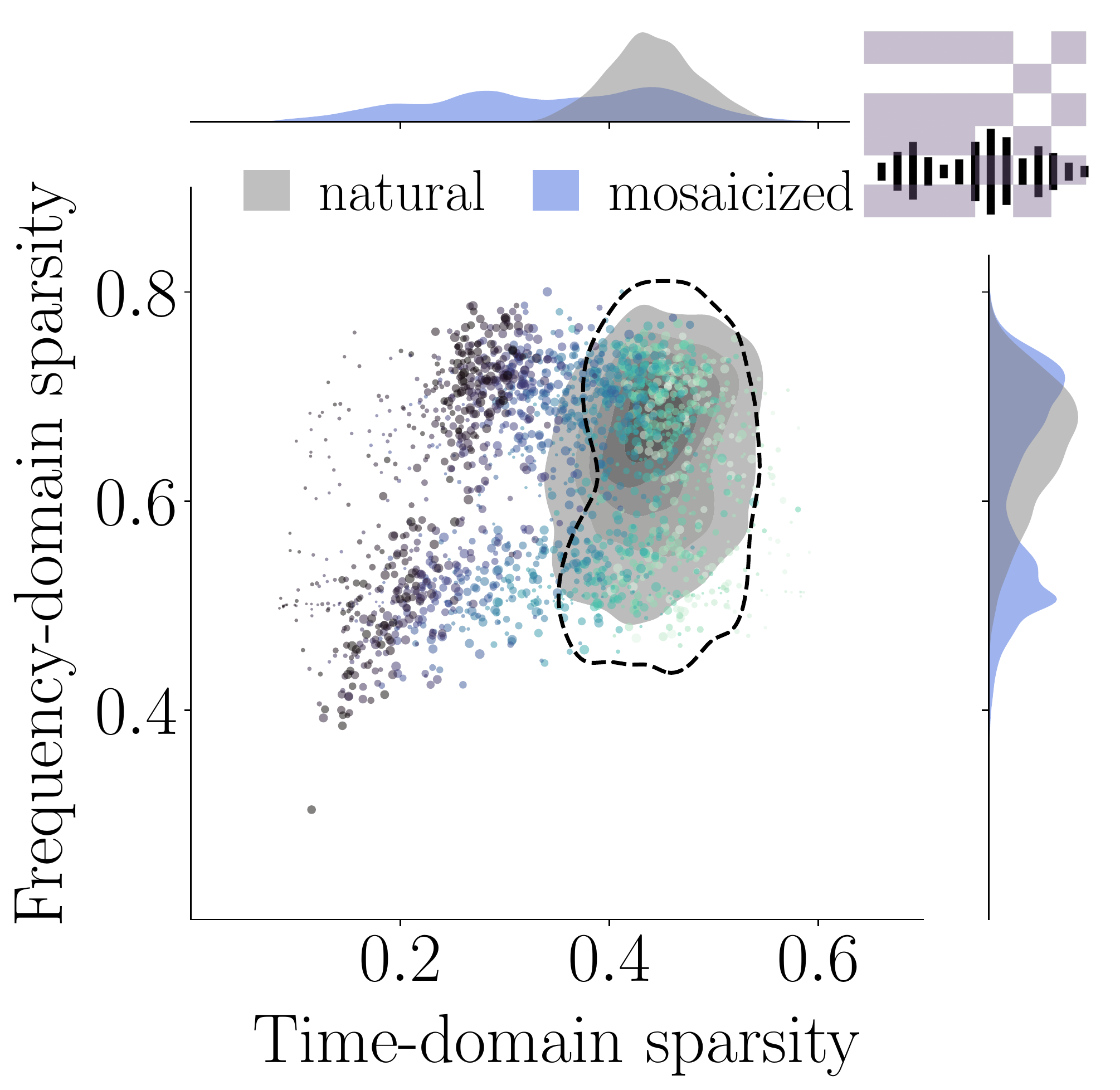}
            \vspace*{2mm}
        \end{subfigure}
        
        \begin{subfigure}[t]{0.32\textwidth}
            \centering
            \subcaption{}
            \includegraphics[scale=\FIGURESCALE]{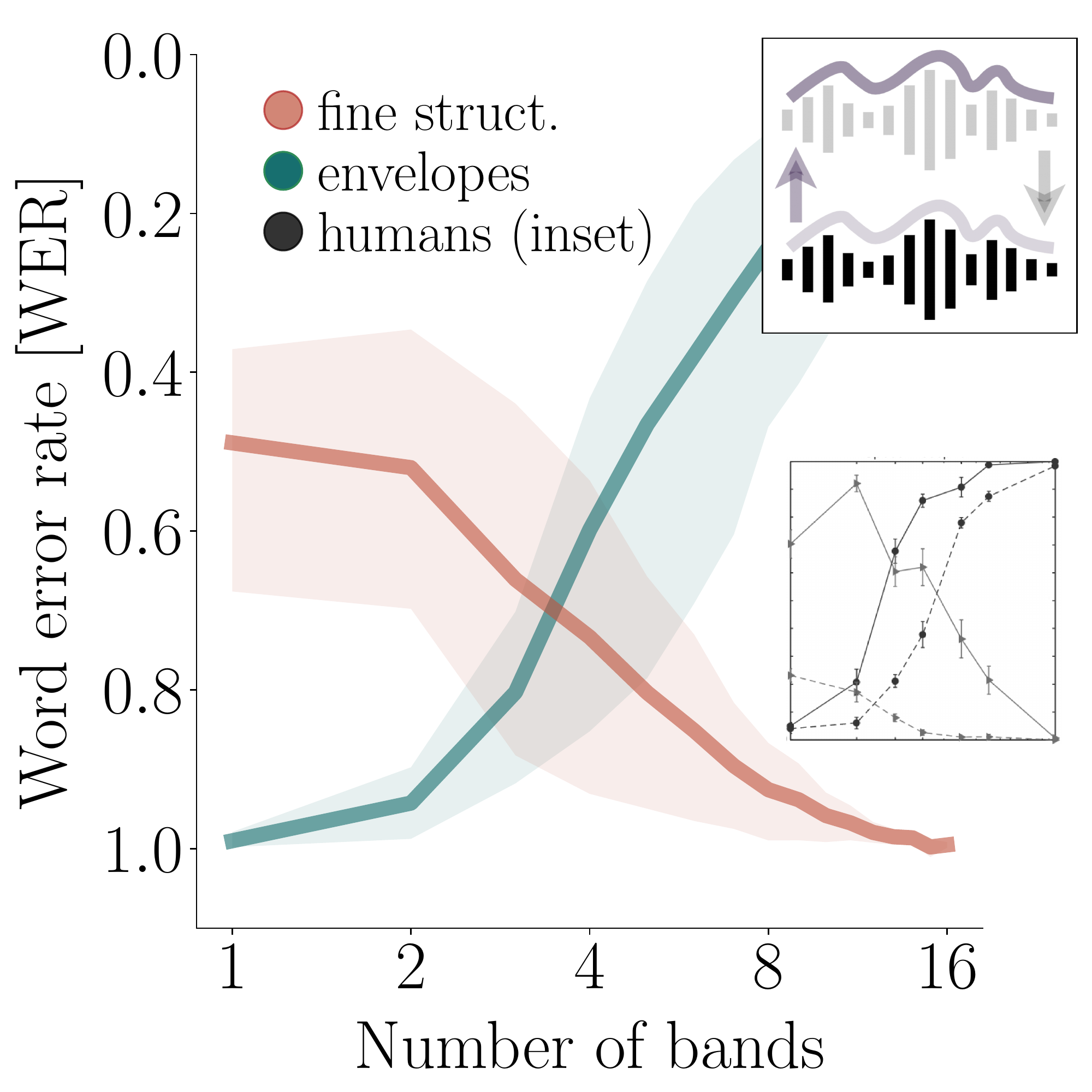}
        \end{subfigure}
            \hfill
        \begin{subfigure}[t]{0.32\textwidth}
            \centering
            \subcaption{}
            \includegraphics[scale=\FIGURESCALE]{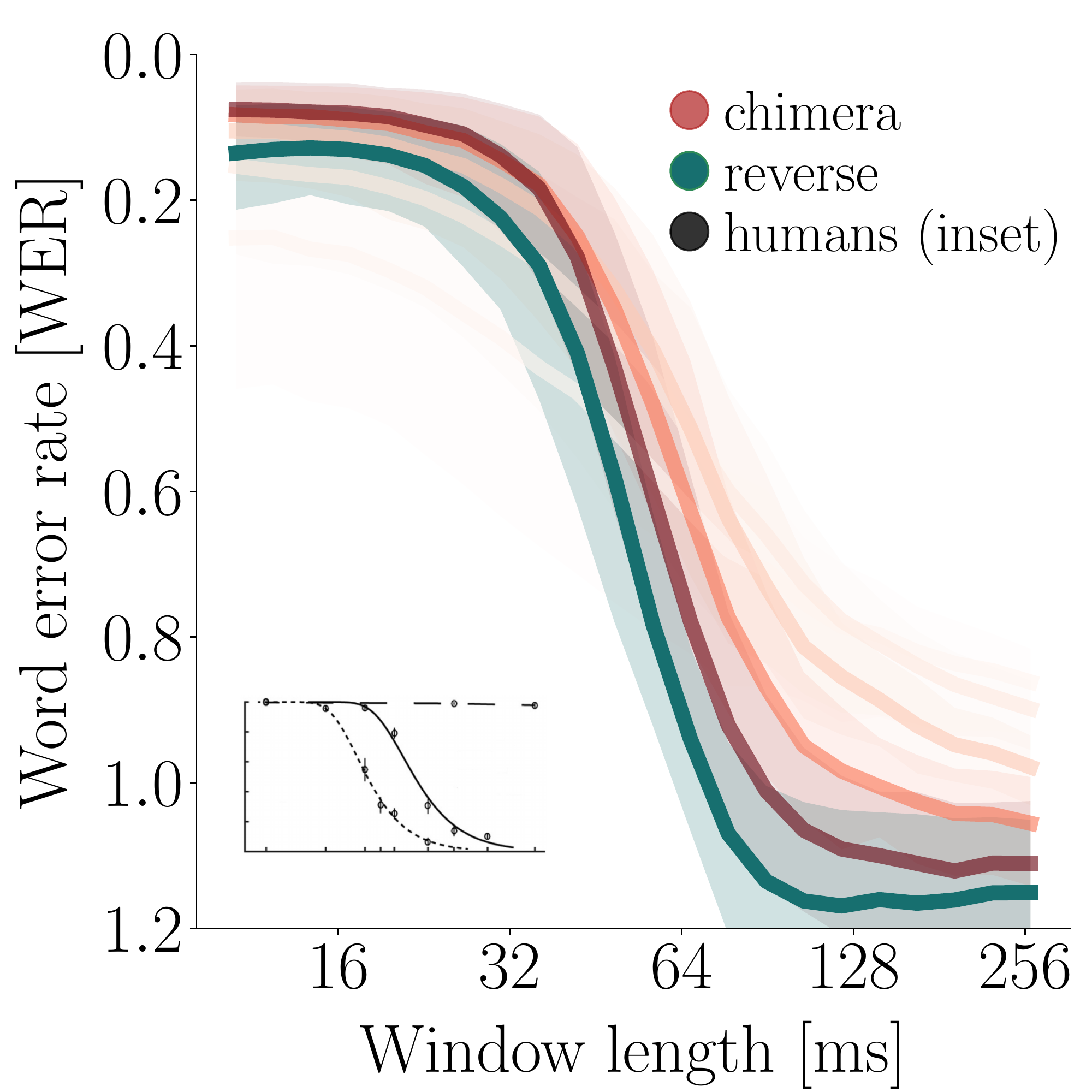}
        \end{subfigure}
            \hfill
        \begin{subfigure}[t]{0.32\textwidth}
            \centering
            \subcaption{}
            \includegraphics[scale=\FIGURESCALE]{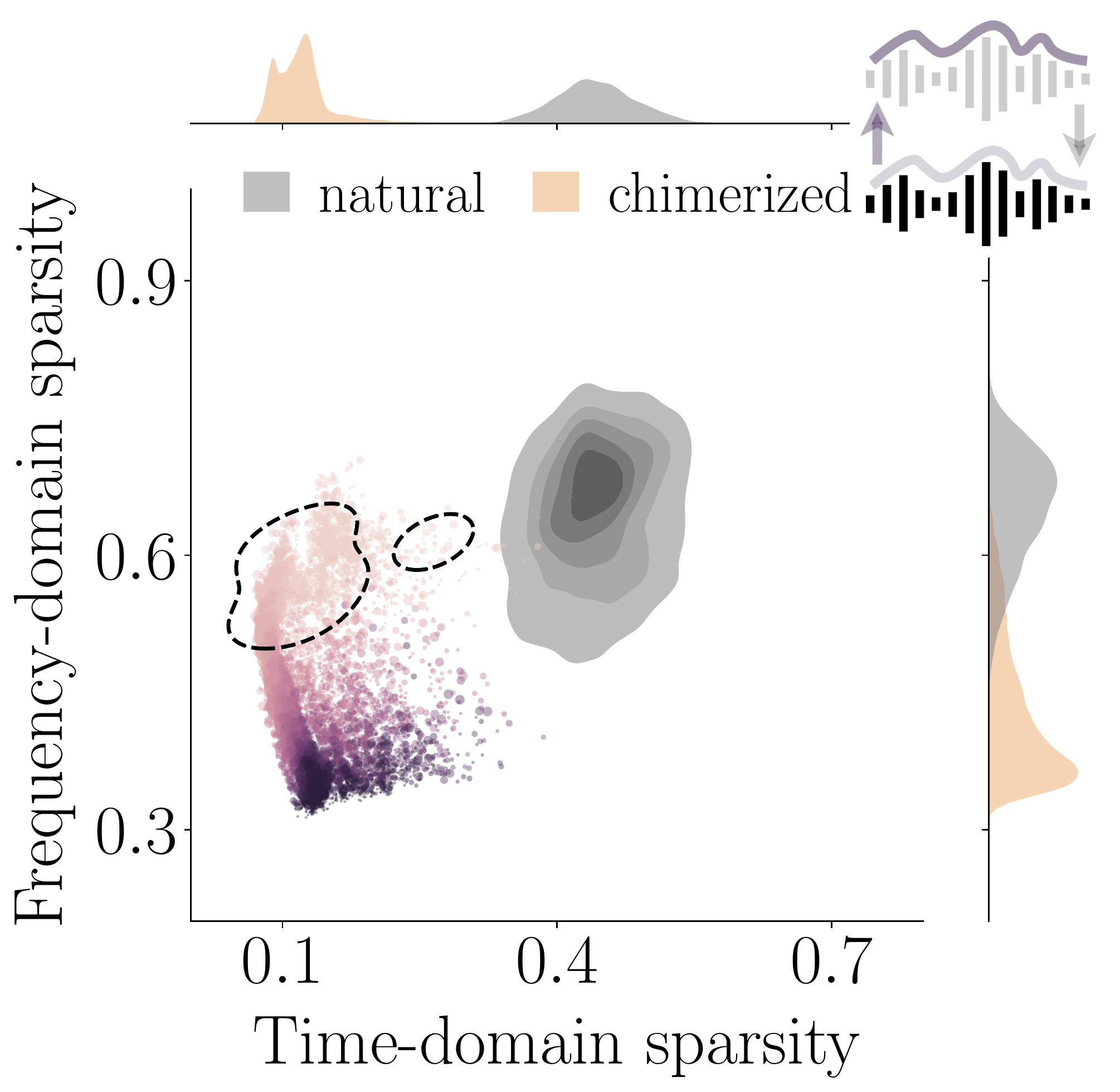}
        \end{subfigure}
        
    \end{center}
\caption{Mosaicized and chimerized speech reveal relatively similar reliance on subband envelopes and fine structure across timescales in humans and machines. We plot performance (WER; left-hand panels) as a function of either window length or number of bands (shade indicates 95\% CI summarizing similar performance across all models; all insets show human performance with x/y-axis ranges comparable to the corresponding main graphs). Speech mosaics (A) with different temporal bin widths (increasing spectral widths shown in lighter shades of red) elicit a uniform advantage relative to multiple-timescale reversal (inset adapted from \citeNP{nakajima2018TemporalResolutionNeededfhn} shows performance for mosaic speech in squares and locally time-reversed speech in triangles, the latter exhibiting a steeper decline). Speech-noise chimeras (C) reveal human-like performance modulations as a function of the number of bands used for synthesis and whether speech information is present in the envelopes or the fine structure (inset adapted from \citeNP{smith2002ChimaericSoundsRevealn} shows increasing performance for envelope in circles and decreasing performance for fine structure in triangles; solid lines represent the relevant speech-noise chimeras). A further time reversal manipulation selectively on the subband envelopes (D; shades of red represent number of bands), preserving speech fine structure, shows a systematic relation to time-domain reversal, as seen in humans (inset adapted from \citeNP{teng2019SpeechFineStructuren} shows performance on time-domain reversal [dotted line] declines earlier than envelope reversal [solid line]). The effect of the manipulations on the input distributions is visualized with hues and sizes representing synthesis parameters B: window length, E: number of bands; dashed contour shows region of 85-100\% model performance)}
\label{figure:chimeramosaic}
\end{figure*}

Since it is natural to think of the family of experiments conducted here  as affecting the distribution of signal energy (in time and frequency) in proportion to the magnitude of the synthesis parameters (see below and Fig. \ref{figure:experiments}), we accordingly use sparsity as a summary statistic. 
We do this with descriptive aims, as it allows us to (a) visualize an interpretable, low-dimensional representation of the input, (b) unify synthesis procedures traditionally considered separate, and (c) reason about out-of-distribution robustness.
To examine how the different speech synthesis techniques relate to each other, we quantify and summarize their effect on the distribution over the input space: we compute the sparsity \cite{hurley2009ComparingMeasuresSparsityacm} of natural and experimental signals in the time and frequency domains. 
A high sparsity representation of a signal contains a small number of its coefficients (under some encoding) accounting for most of the signal's energy. 
We observe that this measure is reliably modulated by our synthesis procedures and experimental parameters, which makes it a useful summary representation of the resulting input statistics. 
We visualize the joint distributions of both synthetic and natural speech samples and find that the family of manipulations approximately fills the space (see Fig. \ref{figure:gini} for a schematic summary). 
Natural speech sits roughly at the center, with the extremes in this space representing regions of canonical non-speech signals like noise, clicks, simple tones, and beeps. 
The magnitude of the experimental manipulations relates to how much synthetic samples are pushed away from natural speech in various directions. 
In the next sections we will present similar graphs alongside the main results, for each experiment separately, to aid in the description of the data.
As we detail in the experiments below, we observe that top performance (80-100\%) on perturbed synthetic speech includes limited regions outside the natural distribution where both humans and machines exhibit inherent robustness (see Fig. \ref{figure:convergence}-\ref{figure:nonrobustness} right-hand panels for individual experiment distributions).
In sum, the family of perturbations considered here are naturally described as spanning the space of sparsity, and can parametrically drive speech stimuli outside the training distribution, where machines and humans exhibit some generalization.

\subsection{Convergent robustness:  artificial systems exhibit humanlike multi-scale invariances to classical perturbations}

We find that machines display qualitatively similar performance patterns to humans in classical experiments where the temporal and spectral granularities, and the perturbations themselves, are manipulated (Fig. \ref{figure:convergence}-\ref{figure:chimeramosaic}). We summarize the findings next.

\textit{Shuffling}, destroys information to a greater extent than, for instance, reversal of the time-domain samples, as it affects the local spectrum.
The manipulation pushes speech towards a region of reduced spectral, and, eventually, temporal sparsity (Fig. \ref{figure:convergence}B).
Consequently, humans show a more dramatic decline with increasing temporal extent \cite{gotoh2017EffectPermutationsTimetjotasoa}.
We observe the same effect in machines (Fig. \ref{figure:convergence}A). 
Performance declines steadily with increasing window size until speech is rendered unrecognizable at around the 2-ms timescale. All models show this basic pattern and cutoff, although with varying rates of decline.

\textit{Reversal}, which affects the temporal order but preserves the local magnitude spectrum — leaving the sparsity statistics largely untouched (Fig. \ref{figure:convergence}D), produces a complicated performance contour in humans \cite{gotoh2017EffectPermutationsTimetjotasoa}.
Perfect performance for window sizes between 5 and 50 ms, and even partial intelligibility for those exceeding 100 ms is readily achieved by humans even though speech sounds carry defining features evolving rapidly within the reversal window. We find that this timescale-specific resistance to reversal \cite{saberi1999CognitiveRestorationReversedn} is closely traced — with increasing precision as more accurate estimates are obtained \cite{ueda2017IntelligibilityLocallyTimereversedsr} — by automatic speech recognition systems (Fig. \ref{figure:convergence}C).

\textit{Time warping} alters the duration of the signal without affecting the spectral content. Similar to size in vision, a system confronted with a time warped sound needs to handle an 'object' that has been rescaled. Humans can cope with stretched or compressed speech with decreasing performance up to a factor of 3-4, with a faster decline for compression \cite{fu2001RecognitionTimedistortedSentencesjasa}. Stretching and compression manifest in the input space as translation in the time-domain sparsity axis in opposite directions (Fig. \ref{figure:convergence}F). We find that neural network performance follows the U-shaped curve found in humans and exhibits the characteristic asymmetry as well (Fig. \ref{figure:convergence}E).
Performance is worst when the warp factor is either 4.0 (compression) or 0.25 (stretching) and it shows a steeper ascent when decreasing compression than when decreasing stretching. The best performance is achieved, as expected, when the warp factor is 1.0 (no compression or stretching, i.e., the natural signal).

\textit{Mosaic sounds} are analogous to pixelated images and therefore better suited than reversed speech to probe the resolution the system needs for downstream tasks \cite{nakajima2018TemporalResolutionNeededfhn}. Values within bins in the time-frequency canvas are pooled such that the representation is 'pixelated'. The size of the bins is manipulated to affect the available resolution. This corresponds to a decrease in sparsity that scales with the spectrotemporal bin size (Fig. \ref{figure:chimeramosaic}B). When the temporal resolution of the auditory system is probed in this way at multiple scales, we find that, as seen in humans, a systematic advantage emerges over the locally reversed counterpart in ANNs (Fig. \ref{figure:chimeramosaic}A).

\textit{Chimaeric sounds} factor the signal into the product of sub-band envelopes and temporal fine structure to combine one and the other component extracted from different sounds. Although the importance of the envelopes has been emphasized \cite{shannon1995SpeechRecognitionPrimarilysa}, recent experiments suggest that this may only be part of the mechanism, with the fine structure having a unique contribution to speech intelligibility \cite{teng2019SpeechFineStructuren}. Speech-noise chimeras can be constructed such that task-related information is present in the envelopes or the fine structure only \cite{smith2002ChimaericSoundsRevealn}. We observe that fine-structure speech shows up as less sparse in the time-domain due to the removal of envelope information, and its frequency-domain sparsity is modulated by the number of bands (Fig. \ref{figure:chimeramosaic}E). Both humans and machines show a characteristic sensitivity to the number of bands used for synthesis: performance over the entire range is boosted or attenuated depending on whether information is present in the envelopes or the fine structure (Fig. \ref{figure:chimeramosaic}C). An additional effect concerns the perceptual advantage of locally reversed speech at the level of sub-band envelopes over both the time-domain waveform reversal and the speech-noise chimera with reversed envelopes \cite{teng2019SpeechFineStructuren}. We find that models, too, exhibit this uniform advantage (Fig. \ref{figure:chimeramosaic}D).
Performance is best when the reversal timescale is roughly less than 50 ms and then rapidly declines and plateaus after the 100-ms timescale where speech is unrecognizable. Following this general trend, the speech-noise chimeras produce the least resilient performance.

\subsection{Divergent robustness:  multi-scale interruptions reveal differential humanlikeness among machines}

\begin{figure*}[ht!]
    \begin{center}
        \centering
        
        \begin{subfigure}[t]{\SUBFIGUREWIDTH\textwidth}
            \centering
            \subcaption{}
            \includegraphics[scale=\FIGURESCALE]{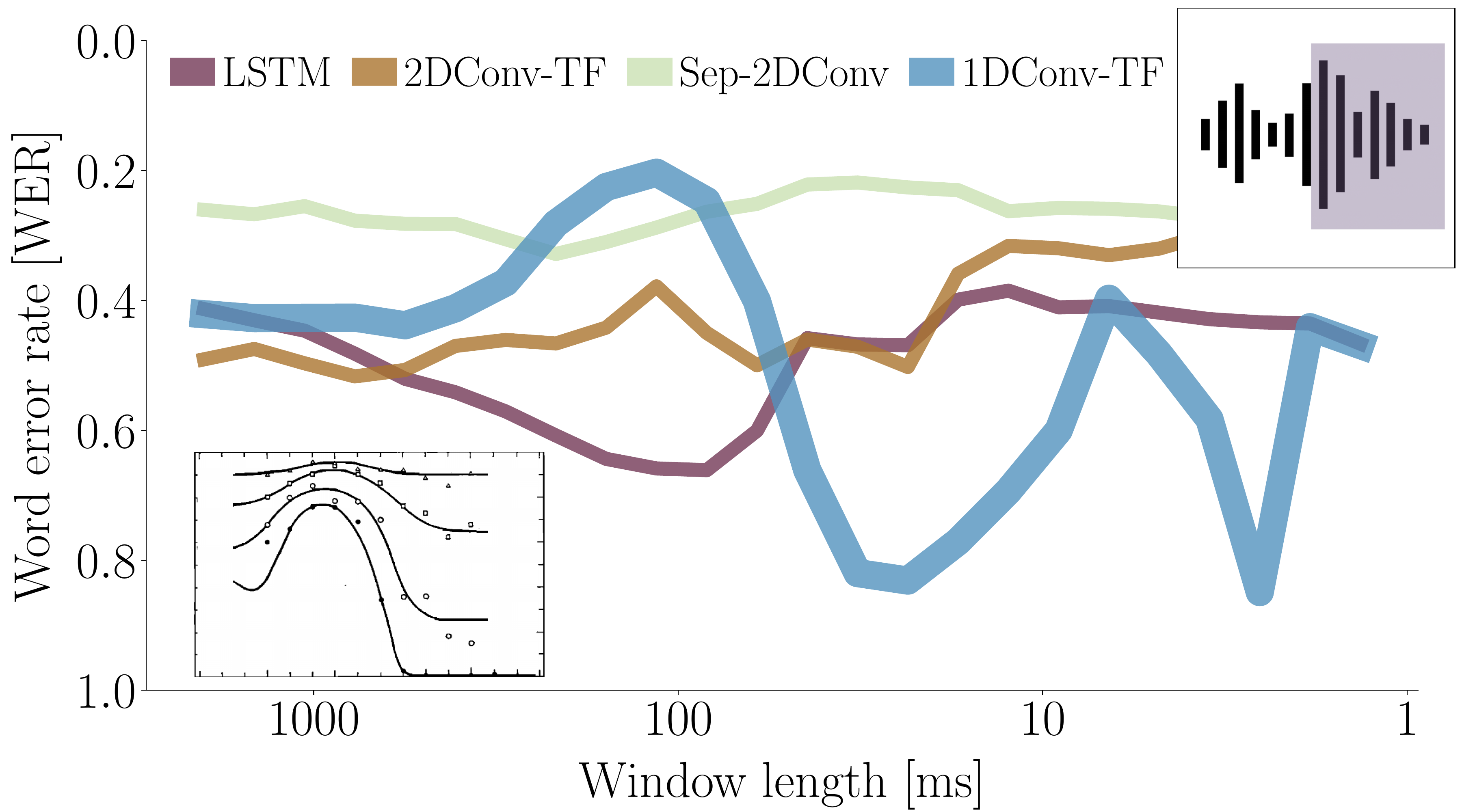}
        \end{subfigure}
            \hfill
        \begin{subfigure}[t]{\SUBFIGUREWIDTH\textwidth}
            \centering
            \subcaption{}
            \includegraphics[scale=\FIGURESCALE]{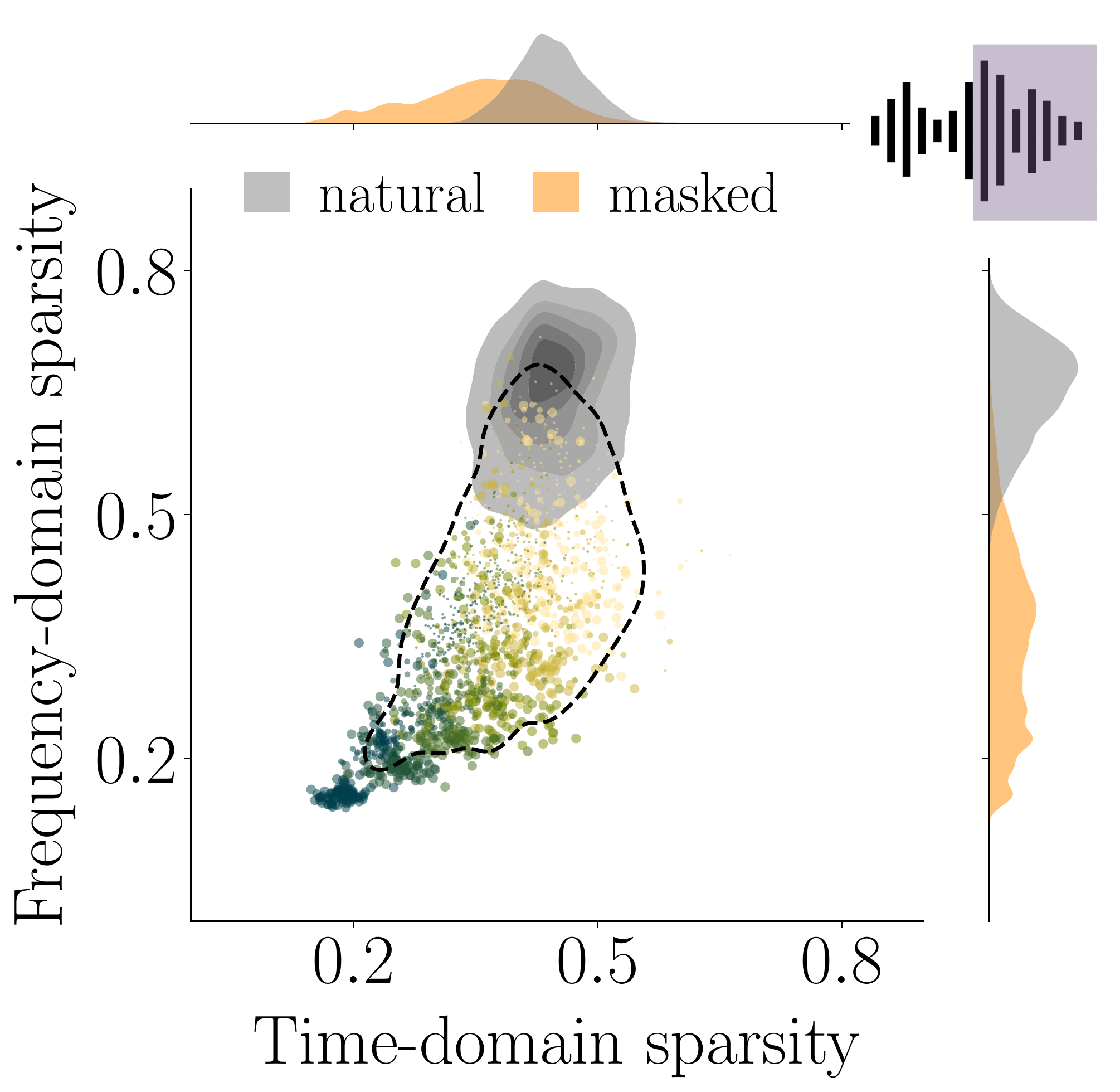}
        \end{subfigure}
        
        \vfill
        
        \begin{subfigure}[t]{\SUBFIGUREWIDTH\textwidth}
            \centering
            \subcaption{}
            \includegraphics[scale=\FIGURESCALE]{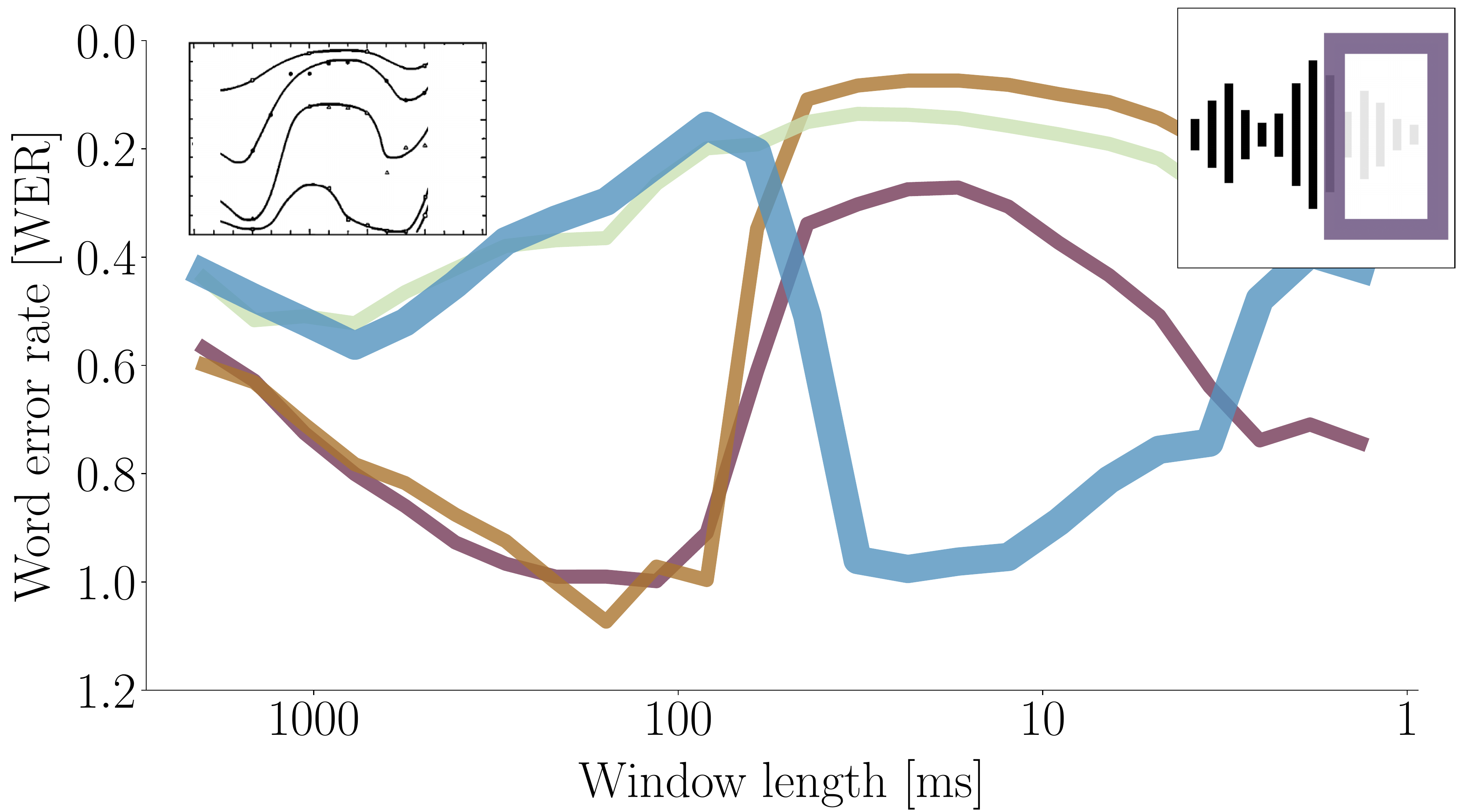}
        \end{subfigure}
            \hfill
        \begin{subfigure}[t]{\SUBFIGUREWIDTH\textwidth}
            \centering
            \subcaption{}
            \includegraphics[scale=\FIGURESCALE]{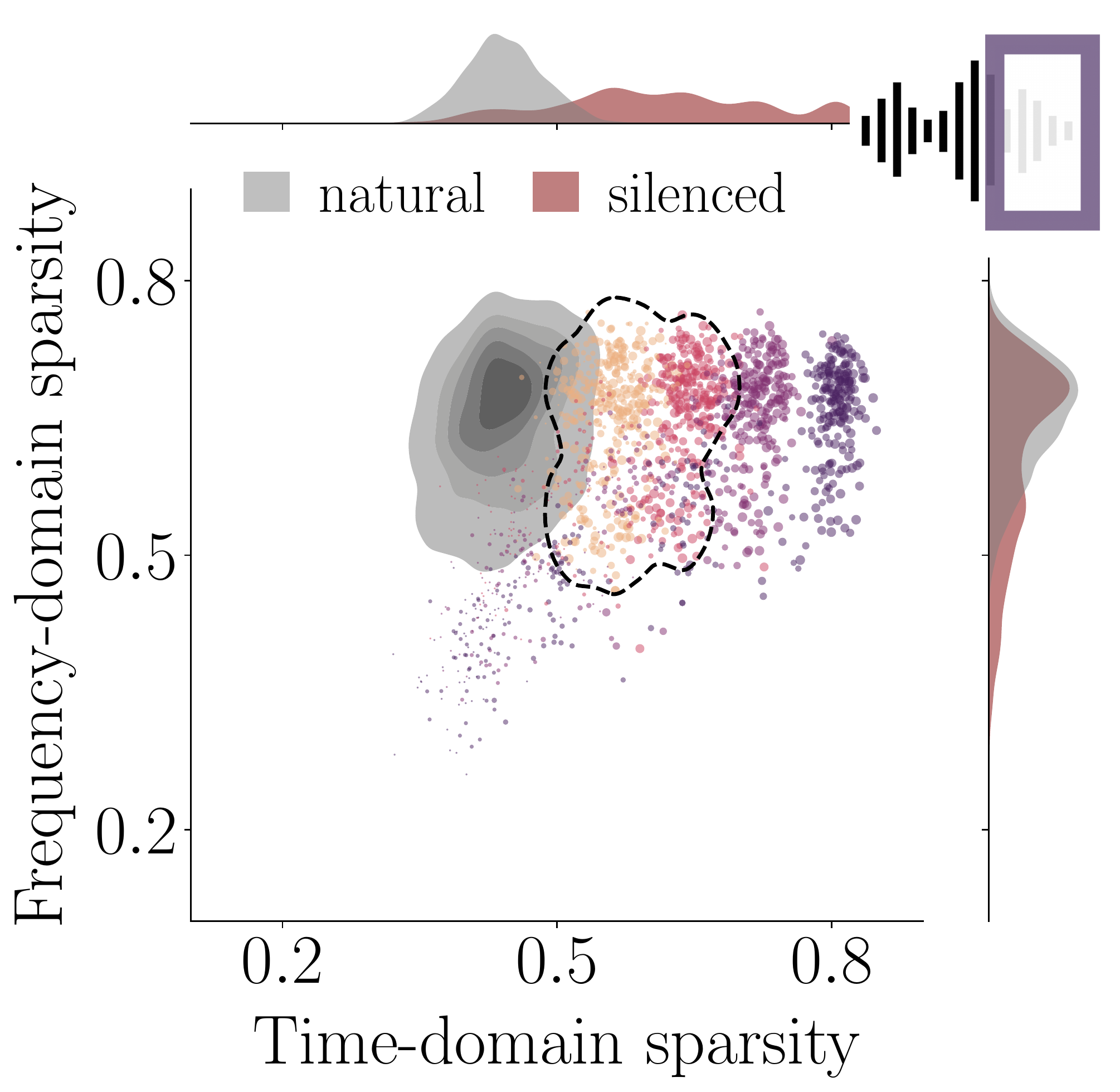}
        \end{subfigure}
        
    \end{center}
\caption{
Multiple-timescale masking and silencing reveals heterogeneity in model predictions (all insets show human performance with x/y-axis ranges comparable to the corresponding main graphs).
We plot performance (WER; left-hand panels) as a function of perturbation timescale (window length in ms).
(A) Masking experiment (inset adapted from \citeNP{miller1950IntelligibilityInterruptedSpeech} shows human performance for various signal-to-noise ratios). 
Individual model performance is shown for a fraction of 0.5 and SNR of -9 db. 
(C) Silencing experiment (inset adapted from \citeNP{miller1950IntelligibilityInterruptedSpeech} shows human performance contours for various silence fractions). 
Individual model performance (color coding on panel A) is shown for a fraction of 0.5 for succinctness.
In both experiments, the transformer architecture with waveform input qualitatively shows the most human-like perceptual behavior. 
The effect of the manipulations on the input distributions (right-hand panels) is visualized with hues and sizes representing synthesis parameters (B: window length, mask fraction, D: window length, silence fraction, respectively; dashed contour shows region of 85-100\% model performance)}
\label{figure:divergence}
\end{figure*}

\textit{Speech interruptions} perturb a fraction of the windows at a given timescale with either silence or a noise mask.
With this manipulation, the system is presented with 'glimpses' of the input signal.
The redundancies in the speech signal are such that at various interruption frequencies, for example between 10 and 100 ms window size, humans show good performance even though a substantial fraction of the input has been perturbed or eliminated \cite{miller1950IntelligibilityInterruptedSpeech}.
Mask interruptions corrupt a fraction of the signal by adding noise.
This shifts the speech samples mainly towards regions of decreasing spectral sparsity (Fig. \ref{figure:divergence}B).
Interruptions of silence, on the other hand, zero out a fraction of the signal, effectively removing all the information in it.
As a consequence, speech samples become increasingly temporally sparse (Fig. \ref{figure:divergence}D).
We find that models exhibit idiosyncratic performance patterns across timescales such that they pairwise agree to different extents depending on the perturbation window size.
Humans, as well as some of the models we tested, exhibit a perceptual profile where obstructions (mask or silence) at large timescales produce moderately bad performance, later recover almost completely at intermediate timescales, they achieve their worst performance at moderately short timescales, and finally slightly improve at the smallest timescales.
As the masking window size decreases from 1000 ms to 100 ms some models' performance declines to their worst and then quickly recover such that they achieve their best at around 50 ms and shorter timescales.
On the other hand, a recent transformer architecture with a waveform front end, pretrained using self-supervision, shows an overall better qualitative match to human performance, although quantitative differences are still apparent in all cases (Fig. \ref{figure:divergence}A,C).

\subsection{Nonrobustness:  machines fail to exhibit humanlike performance profiles in response to repackaging}

\begin{figure*}[!ht]
    \begin{center}
        \centering
        \begin{subfigure}[t]{\SUBFIGUREWIDTH\textwidth}
            \centering
            \subcaption{}
            \includegraphics[scale=\FIGURESCALE]{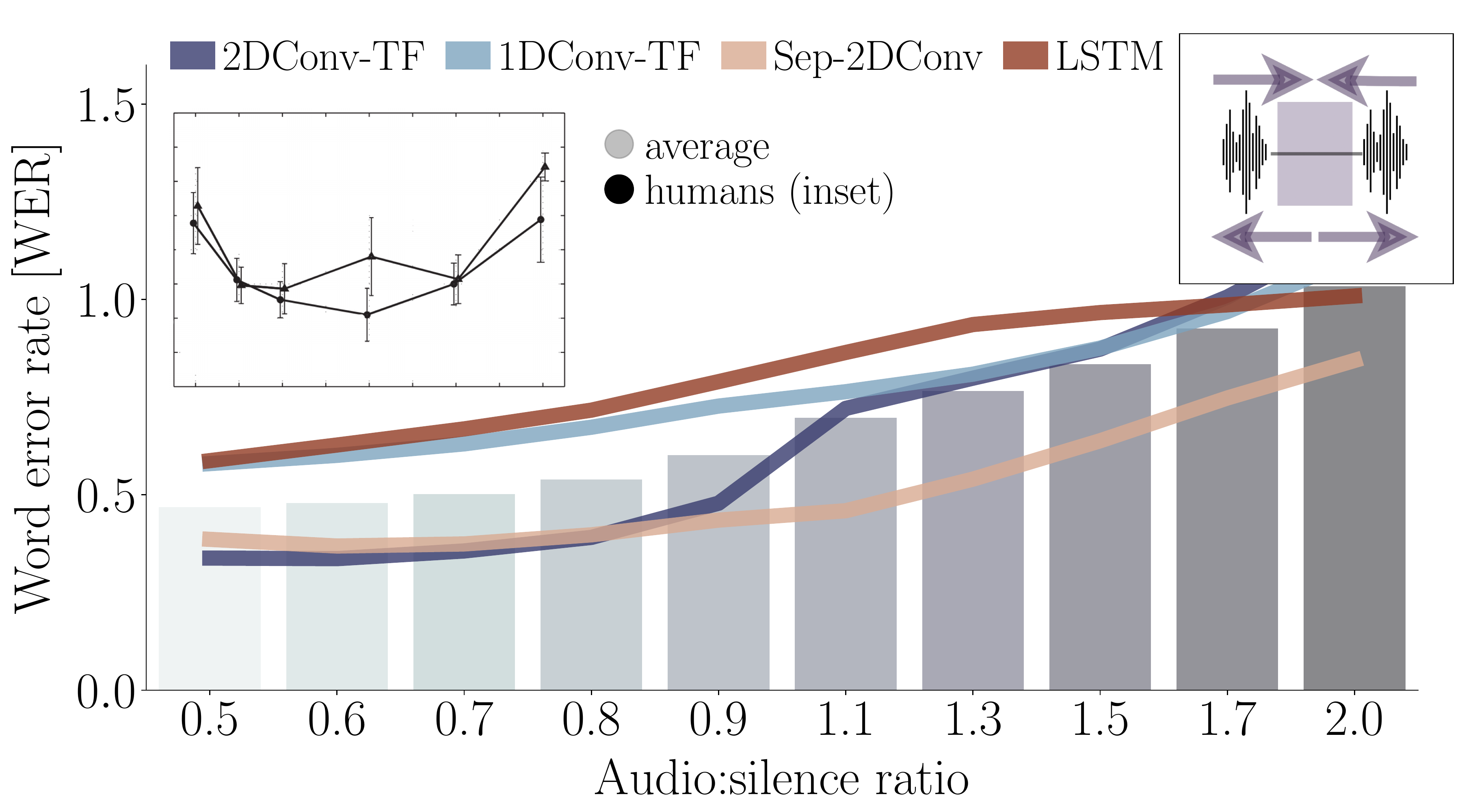}
        \end{subfigure}
        \hfill
        \begin{subfigure}[t]{\SUBFIGUREWIDTH\textwidth}
            \centering
            \subcaption{}
            \includegraphics[scale=\FIGURESCALE]{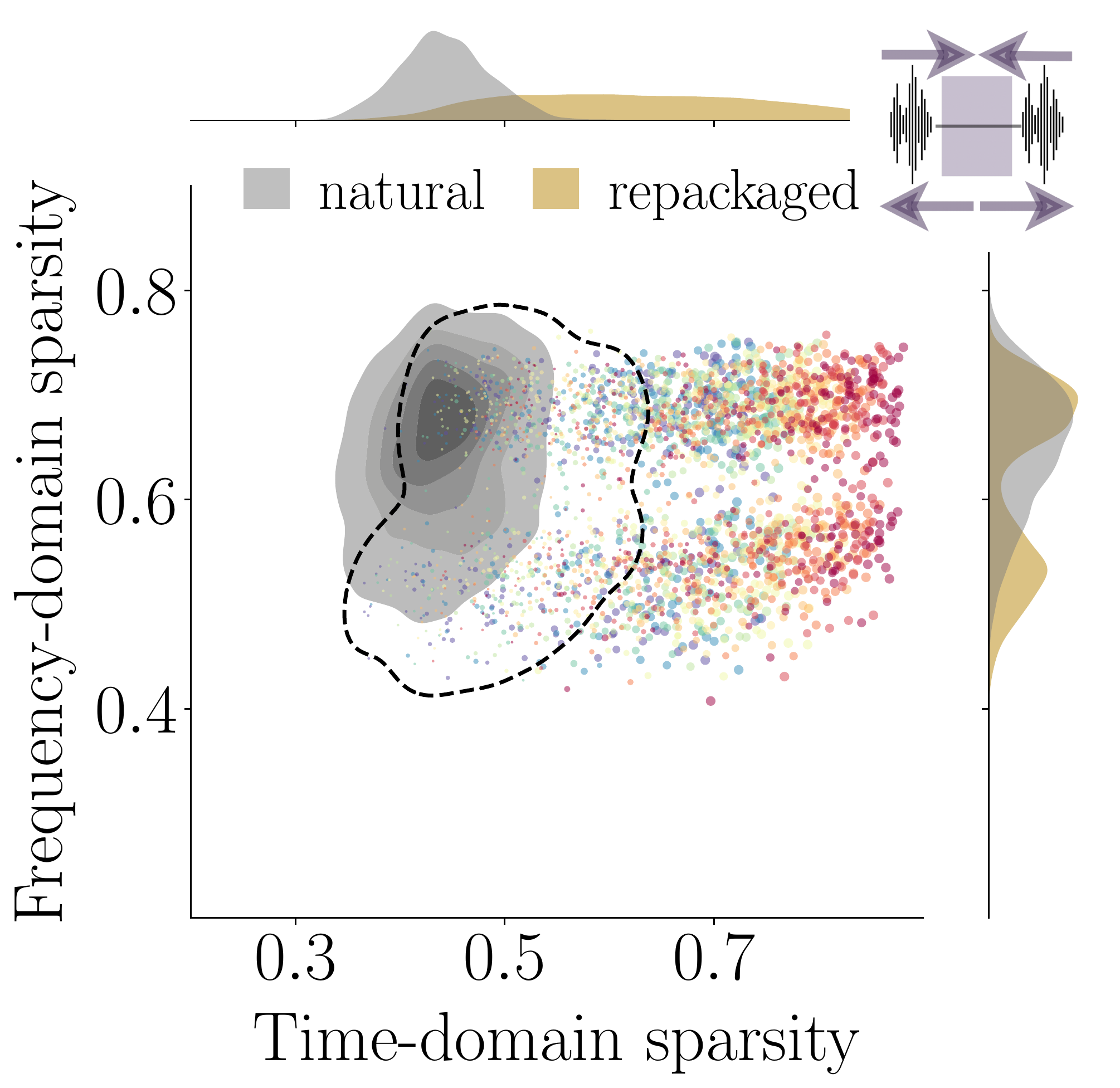}
        \end{subfigure}
    \end{center}
\caption{None of the architectures predict recovery of performance with repackaging as seen in humans (A; inset adapted from \citeNP{ghitza2009PossibleRoleBrainp} shows the canonical U-shaped human performance pattern, with x/y-axis ranges comparable to the main graph; solid lines with circle markers represent the relevant manipulation with insertions of silence. We plot performance (WER; left-hand panel) on compressed speech (by a factor of 2) as a function of audio:silence ratio parameterizing the insertion of silence. Note here the y-axis is reversed, with lower error towards the origin). Although there is some robustness to compression outside the natural distribution, performance worsens steadily as the insertion length increases. The effect of compressed audio with insertions of silence on sparsity is visualized with hues and sizes representing synthesis parameters (B; dashed contour shows region of 85-100\% model performance).}
\label{figure:nonrobustness}
\end{figure*}

\textit{Repackaging} combines different aspects of the previous audio manipulations — time warping, multiple-timescale windowing, and insertions of silence — to reallocate, as opposed to remove or corrupt, speech information in time. Repackaged speech therefore can be made more temporally sparse (Fig. \ref{figure:nonrobustness}B) without losing information.
As we have shown above, the performance of both humans and machines degrades with increasing temporal compression. Here we focus further on a key finding: when perceiving compressed speech humans benefit from repackaging \cite{ghitza2009PossibleRoleBrainp}.
Insertions of silence, roughly up to the amount necessary to compensate for the compression, recovers performance dramatically — an effect that has been replicated and further characterized numerous times \cite{ghitza2009PossibleRoleBrainp,ghitza2012RoleThetaDrivenSyllabicfp,ghitza2014BehavioralEvidenceRolefp,bosker2018EntrainedThetaOscillationslcn,penn2018PossibleRoleBrainjasa,ramus2021IntelligibilityTemporallyPackaged}.
We find that, across the entire space of experimental parameters, machines fail to show any such recovery (Fig. \ref{figure:nonrobustness}A).
The canonical performance profile in humans shows the worst performance when the signal is compressed by a factor approaching 3 and no silence is inserted. 
As the amount of silence inserted compensates for the extent lost due to temporal compression, the performance improves. 
After that, it declines, producing a characteristic U shape.
The systems tested here, on the other hand, show bad performance with heavily compressed speech which simply worsens with increasing insertions of silence and shows no inflection when the insertion length precisely compensates for the temporal compression.

\section{Discussion}

In this paper we considered the possibility that engineering solutions to artificial audition might qualitatively converge, in more ways than merely performance level, with those implemented in human brains. If the task is too simple, then it is conceivable that many qualitatively different solutions can in principle be possible. In this case, convergence of performance between humans and engineered systems would not be surprising. On the other hand, convergence of algorithmic solutions would be. If the constraints on the task become more nuanced, however, then any system learning to solve the task would be forced into a narrower space of possible algorithmic solutions. In this latter case, similar performance levels might be suggestive of similar algorithmic solutions. Here we set out to investigate whether this might be the current scenario regarding human speech perception and neural networks for automatic speech recognition.

In a set of studies we had high-performing speech-to-text neural networks perform theoretically-driven tasks while sweeping through the parameter space of foundational speech experiments.
We additionally explored how the audio perturbations in each experiment relate to their unpurturbed counterparts and to canonical audio signals.
We found that a subset of multi-scale perturbations including reversal, shuffling, time warping, chimerizing and mosacizing yield performance curves and effects that are similar amongst models and to some extent aligned with those of humans. More destructive perturbations such as masking and silencing reveal performance patterns where models differ from each other and from humans. The most informative outcome we observed comes from the repackaging experiment, whereby all models resemble each other closely while systematically failing to capture human performance. This finding highlights a set of possible endogenous mechanisms currently absent from state-of-the-art neural network models. We focus here on the broad qualitative trends that are informative for theory and model development as we discuss the implications for the reverse-engineering and (more speculatively) the forward-engineering of hearing.

We found that several classical phenomena in speech perception are well-predicted by high-performing models.
These comprise the performance curves across scales in response to reversed \cite{saberi1999CognitiveRestorationReversedn}, shuffled \cite{gotoh2017EffectPermutationsTimetjotasoa}, time-warped \cite{fu2001RecognitionTimedistortedSentencesjasa}, mosacized \cite{nakajima2018TemporalResolutionNeededfhn}, and chimerized \cite{smith2002ChimaericSoundsRevealn} speech. 
Humans and machines perform well in these non-ecological conditions at qualitatively similar scales, and this emerges simply as a result of training on the downstream task of recognition.
This need not have been the case, for example, if different solutions to the problem are possible (e.g., equally predictive cues) and systems have various inductive biases that push towards them differently.
Overall, these similarities could be interpreted as a form of shared out-of-distribution robustness: neither humans nor machines need any specific training to achieve it.
These effects do not correspond to the perturbations having no measurable effect whatsoever, as they are known to lie well above the detection threshold; stimuli appear unnatural even to untrained listeners and the results generally agree with foundational studies (e.g., \citeNP{shannon1995SpeechRecognitionPrimarilysa}).
Broad agreement between different architectures for multiple-spectrotemporal-scale manipulations, as we have found, would tentatively suggest that the problem of speech recognition offers enough constraints such that humans and artificial systems naturally converge in high-performance regimes.
This is the prevailing view behind studies predicting brain activity using these kinds of models (e.g., \citeNP{milletInductiveBiasesPretraining2021}): that high-performing networks settle on hyperparameter regions which, although chosen for engineering reasons, turn out to be human-like in some relevant way (e.g., having similar receptive field sizes).

However, we observe some marked differences emerging among artificial systems and between these and humans when the signal is perturbed more aggressively. 
These comprise the masking and silencing manipulations \cite{miller1950IntelligibilityInterruptedSpeech}, where the performance profiles vary more widely.
The perturbations we deploy are not natural but they have been designed to probe attributes of perception that the developing auditory system must acquire as it is confronted with natural signals, such as resilience to temporal distortions due to reverberation and various forms of masking.
The possible reasons for differences between the models themselves are of secondary importance here as we are specifically concerned with their ability or not to capture qualitative human behavioral patterns.
Although there might be a way to reconcile these diverse performance patterns by altering minor parameters in the architectures, our work together with a parallel effort using different methods \cite{weerts2021PsychometricsAutomaticSpeechb} highlights a more fundamental difficulty of these architectures to perform well in the presence of noise.
Weerts et al. \citeyear{weerts2021PsychometricsAutomaticSpeechb} compared 3 artificial systems with human performance using a test battery of psychometric experiments (e.g., spectral invariance, peak and center clipping, spectral and temporal modulations, target periodicity, competing talker backgrounds, and masker modulations and periodicity) to measure the importance of various auditory cues in sentence- or word-based speech recognition. They find that systems display similarities and differences in terms of what features they are tuned to (e.g., spectral vs. temporal modulations, and the use of temporal fine structure).
As in our work, the self-supervised CNN-Transformer model exhibited a relatively greater similarity to humans, which follows a recent trend in vision \cite{tuli2021AreConvolutionalNeuralac}. 

Both these similarities and differences have alternative interpretations.
With regards to the dissimilarities, it could be argued that the performance patterns point to differently tuned parameters of similar mechanisms (e.g., different effective receptive field sizes of architecturally similar systems), or alternatively, to more important mechanistic differences.  With regards to the similarities, the results could be a consequence of how information is distributed in the input signal (i.e., where in the signal and envelope spectra information is carried), and as such, not provide compelling evidence that these models processed signals in a human-like way.
By visual analogy, if image content was consistently concentrated in certain locations in the canvas, perturbations applied systematically and selectively across the canvas would affect similarly any systems that make use of such information (i.e., produce similarly complicated performance curves).
This certainly tells us about the way task-related information is distributed in the signal and that high-performing problem solutions are constrained to the set that exploit such information, but it does not provide much mechanistic insight otherwise.
On the other hand, these similarities may reflect important aspects of convergence between human and machine solutions.
Therefore, although this class of findings is informative in many ways, the outcomes do not point unambiguously to mechanistic differences and similarities. 

The repackaging experiments, however, yield consistent and unambiguous failures that allow stronger conclusions to be drawn.
The perception of temporally repackaged speech \cite{ghitza2009PossibleRoleBrainp,ghitza2012RoleThetaDrivenSyllabicfp} is a scenario where the similarity between neural network models and their substantial deviation from human performance is remarkably consistent.
Our repackaging experiments demonstrate a systematic failure of all models to recover perceptual performance in the specific conditions that humans naturally do: when the windowed compression of speech is compensated by insertions of silence.
This consistent pattern emerges across diverse models, demonstrating its robustness against substantial architectural variation.
Our simulations cover the whole set of experimental parameter combinations such that we can rule out the presence of the effect even in cases where it would show up in a parameter region away from where experiments in humans have been specifically conducted (e.g., for different compression ratios or window sizes).

The human behavioral profile in response to repackaged speech \cite{ramus2021IntelligibilityTemporallyPackaged, ghitza2009PossibleRoleBrainp} can be interpreted in landmark-based (e.g., 'acoustic edges') and oscillation-based (e.g., theta rhythms) frameworks.
On the former view (e.g., \citeNP{oganianSpeechEnvelopeLandmark2019,hamiltonParallelDistributedEncoding2021}) acoustic cues in the signal envelope increasingly resemble the original as compression is compensated by insertions of silence.
On the latter view \cite{ghitza2009PossibleRoleBrainp,ghitza2012RoleThetaDrivenSyllabicfp,ghitza2014BehavioralEvidenceRolefp}, which has been the subject of further developments regarding neural implementation \cite{poeppel2020SpeechRhythmsTheirnrna,giraudCorticalOscillationsSpeech2012,tengThetaBandOscillations2017,tengThetaGammaBands2019}, insertions of silence enable an alignment with endogenous time constraints embodied by neural oscillations at specific time scales.
A related conceptual framework, which is compatible with both the acoustic landmark and oscillation-based accounts, explains the effect in terms of concurrent multiple-timescale processing \cite{poeppel2008SpeechPerceptionInterfaceptrsb, poeppel2003AnalysisSpeechDifferentsc,tengConcurrentTemporalChannels2017,tengTestingMultiscaleProcessing2016}: the auditory system would elaborate the input signal simultaneously at 2 timescales (roughly, 25-50 and 150-250 ms), and therefore an inherent compensatory strategy when the local information is distorted (e.g., compressed) is to perform the task based on the global information that remains available (e.g., due to insertions of silence).
The important point for present purposes is that all these accounts of repackaged speech involve endogenous mechanisms (e.g., neural excitability cycles) currently absent from state-of-the-art neural network models, with each theoretical proposal attributing model failures to these architectural shortcomings.
These might be crucial for a better account of human audition, and could provide inductive biases for machines that might enable robustness in various real-world auditory environments.
A promising direction, therefore, is incorporating oscillations into the main mechanisms of computational models (e.g., \citeNP{effenbergerBiologyinspiredRecurrentOscillator2022, kaushikMathematicalNeuralProcess2022, tenoeverOscillatingComputationalModel2021}), or otherwise introducing commitments to dynamic temporal structure (e.g., using spiking neural networks; \citeNP{stimbergBrianIntuitiveEfficient2019}) beyond excitability cycles.

A further line of reasoning about the dissimilarities observed in repackaging experiments has to do with the computational complexity of the processes involved \cite{vanrooijParameterizedComplexityCognitive2007}.
Repackaging manipulations have been interpreted as tapping into segmentation \cite{ghitza2009PossibleRoleBrainp, ghitza2014BehavioralEvidenceRolefp} — a subcomputation that has been widely assumed to be computationally hard in a fundamental way (surveyed briefly in \citeNP{adolfiComputationalComplexitySegmentation2022c}; e.g.,  \citeNP{fristonActiveListening2021, poeppel2003AnalysisSpeechDifferentsc, cutlerPerceptionRhythmLanguage1994}).
On this view, any system faced with a problem that involves segmentation as a sub-problem (e.g., speech recognition) would be forced to acquire the (possibly unique) solution that, through exploiting ecological constraints, renders the problem efficiently computable in the restricted case. 
However, contrary to common intuitions, it is possible that segmentation is efficiently computable in the absence of such constraints \cite{adolfiComputationalComplexityPerspective2022a}.
Since it is conceivable segmentation is not a computational bottleneck in this sense, its intrinsic complexity might not be a driving force in pushing different artificial or biological systems to acquire similar solutions to problems involving this subcomputation.
This constitutes, from a theoretical and formal standpoint, a complementary explanation for the qualitative divergence between humans and machines observed in our results.

\subsection{Closing remarks}
Our work and recent independent efforts \cite{weerts2021PsychometricsAutomaticSpeechb} suggest that, despite some predictive accuracy in neuroimaging studies (\citeNP{milletInductiveBiasesPretraining2021,kellTaskOptimizedNeuralNetwork2018, tuckuteManyNotAll2022}; but see \citeNP{thompsonEffectTaskTraining2019}), automatic speech recognition systems and humans diverge substantially in various perceptual domains. 
Our results further suggest that, far from being simply quantitative (e.g., receptive field sizes), these shortcomings are likely qualitative (e.g., lack of flexibility in task performance through exploiting alternative spectrotemporal scales) and would not be solved by such strategies as introducing different training regimens or increasing the models' capacity. 
They would require possibly substantial architectural modifications for meaningful effects such as repackaging to emerge.
The qualitative differences we identify point to possible architectural constraints and improvements, and suggest which regions of experimental space (i.e., which effects) are useful for further model development and comparison.
Since repackaging is where all models systematically resemble each other and clearly fail in capturing human behavior, this effect offers alternative directions for theorizing, computational cognitive modeling, and, more speculatively, potential improvement of engineering solutions.

To develop a deeper understanding of how the models themselves can be made independently more robust, one could implement data augmentation schemes with the perturbations we deployed here. It is conceivable that these could act as proxies for the natural distortions humans encounter in the wild, and therefore help close performance gaps where it is desired for engineering purposes. A related line of research could pursue the comparison of frontend-backbone combinations to evaluate whether particular pairings are effective in some systematic manner in combination with such data augmentation schemes.

More generally, our approach and results showcase how a more active synergy between the cognitive science and engineering of hearing could be mutually beneficial. Historically, there was a close relationship between work in the cognitive sciences, broadly construed, and engineering. Researchers were mindful of both behavioral and neural data in the context of building models (e.g., \citeNP{ghitzaAuditoryNerveRepresentation1986}; Bell Laboratories). Perhaps as a consequence of exclusively quantitative, benchmark-driven development and the recent disproportionate focus on prediction at the expense of explanation (see Bowers et al., 2022, for a review of how this played out in vision), this productive alliance has somewhat diminished in its depth and scope, but the potential gains from a possible reconnection and realignment between disciplines are considerable.


\section{Methods}

\subsection{Framework}
Generalizing from the particular studies examining audition at multiple scales, we build a unified evaluation environment centered around selective transformations at different spectrotemporal granularities
\footnote{Code implementing the analyses and resynthesis methods described here is available at https://tinyurl.com/2e5echc8} 
and their influence on perceptual performance (Fig. \ref{figure:experiments}).
We deliberately shift the focus away from quantitative measures of fit and towards a qualitative assessment (see \citeNP{navarroDevilDeepBlue2019}, for details on the rationale).
Contrary to problematic practices centered on predictive accuracy that have led to misleading conclusions (see \citeNP{bowersDeepProblemsNeural2022} for a thorough review), we focus on assessing whether artificial systems – here treated as stimulus-computable, optimized observer models — qualitatively capture a whole family of idiosyncratic performance patterns in human speech perception that point to the kinds of solutions systems have acquired. 
Our framework situates existing experiments in humans as a subset of the possible simulations, allowing us to exhaustively search for qualitative signatures of human-like performance even when these show up away from the precise original location in experimental space (we show a representative summary of our results throughout).

\subsection{Audio synthesis}

\subsubsection{Multiscale windowing.}Common to all conditions is the windowing of the signal separately at multiple spectral and/or temporal scales. We used a rectangular window function to cut the original speech into frames and faded the concatenated frames to avoid discontinuities (although these turn out to not affect the results). Transformations with known properties (see below) are applied either in the time domain directly or in the time-frequency domain, to each window (i.e., chunk of the signal). The window size is a general parameter that determines the scale selectivity of the manipulations described below. The timescales depend on the experiment and range from a few milliseconds to over one second. The performance of the models under different perturbations is then evaluated separately at various scales.

\subsubsection{Reversal.}
The signal is locally reversed in time \cite{saberi1999CognitiveRestorationReversedn,gotoh2017EffectPermutationsTimetjotasoa}, resulting in frame-wise time-reversed speech. This affects the order of the samples but preserves the local average magnitude spectrum. The performance curve is estimated at 58 timescales on a logarithmic scale ranging from 0.125 to 1200 ms.

\subsubsection{Shuffling.} Audio samples are locally shuffled such that the temporal order within a given window is lost, consequently destroying temporal order at the corresponding scale. This random permutation is more aggressive than reversal in the sense that it does affect the local magnitude spectrum \cite{gotoh2017EffectPermutationsTimetjotasoa}. The performance curve is estimated at 58 timescales on a logarithmic scale ranging from 0.125 to 1200 ms.

\subsubsection{Time warping.}
Signals are temporally compressed or stretched in the time-frequency domain, effectively making speech faster or slower, such that the pitch is unaffected \cite{park19e_interspeech}. The modified short-time Fourier transform is then inverted to obtain the final time-domain, time-warped signal \cite{Perraudin2013griffinLim}. The average magnitude spectrum is approximately invariant whereas the local spectrum is equivariant when compared between equivalent timescales \cite{ghitza2009PossibleRoleBrainp, fu2001RecognitionTimedistortedSentencesjasa}. The performance curve is estimated at 40 parameter values on a logarithmic scale ranging from compression by a factor of 4 to stretching by a factor of 4.

\subsubsection{Chimerism.}
Signals are factored into their envelope and fine structure parts, allowing the resynthesis of chimeras which combine the slow amplitude modulations of one sound with the rapid carriers of another \cite{smith2002ChimaericSoundsRevealn}. Here we combine these features from speech and Gaussian noise. To extract the two components, signals are passed through a bank of band-pass filters modeled after the human cochlea, yielding a spectrogram-like representation in the time-frequency domain called cochleagram. A cochleagram is then a time-frequency decomposition of a sound which shares features of human cochlear processing \cite{glasbergDerivationAuditoryFilter1990}. Using the filter outputs, the analytical signal is computed via the Hilbert transform. Its magnitude is the envelope part, and dividing it out from the analytic signal leaves only the fine structure. The spectral acuity of the synthesis procedure can be varied with the number of bands used to cover the bandwidth of the signal. Multiple timescale manipulations, such as reversal, are directed to either the envelope or fine structure prior to assembling the sound chimera \cite{teng2019SpeechFineStructuren}.  The performance curve is estimated at 32 timescales on a logarithmic scale ranging from 10 to 1200 ms.

\subsubsection{Mosaicism.}
Speech signals can be mosaicized in the time and frequency coordinates, by manipulating the coarse-graining of the time-frequency bins. Similar to a pixelated image, a mosaicized sound will convey a signal whose spectrotemporal resolution has been altered systematically. The procedure is done on the envelope-fine-structure representation before inverting back to the waveform. Two parameters affect the spectral and temporal granularity of the manipulation: the window length in time and in frequency. This yields a grid in the time-frequency domain. The envelope in each cell is averaged and the 'pixelated' cochlear envelopes are used to modulate the fine structure of Gaussian noise. Finally the signal is resynthesized by adding together the modulated sub-bands \cite{nakajima2018TemporalResolutionNeededfhn}. The performance curve is estimated at 32 timescales on a logarithmic scale ranging from 10 to 1200 ms.

\subsubsection{Interruptions.}
A fraction of the within-window signal, which is parametrically varied, is corrupted either with Gaussian noise at different signal-to-noise ratios or by setting the samples to zero \cite{miller1950IntelligibilityInterruptedSpeech}. Sparsity is increased or decreased while preserving the original configuration of the unmasked fraction of the signal.  The performance curve is estimated at 30 timescales on a logarithmic scale ranging from 2 to 2000 ms.

\subsubsection{Repackaging.}
A repackaged signal locally redistributes the original chunks of samples in time \cite{ghitza2009PossibleRoleBrainp,ramus2021IntelligibilityTemporallyPackaged}. Within each window, the signal is temporally compressed without affecting its pitch (see above) and a period of silence is concatenated. The time-compressed signal can alternatively be thought of as a baseline before adding the insertions of silence. Two parameters control the sparsity of the resulting signal: the amount of compression and the length of the inserted silence. Other parameters that mitigate discontinuities, such as additive noise and amplitude ramps, do not affect the results. For a signal that has been compressed by a factor of 2, inserting silence of length equal to 1/2 of the window size will locally redistribute the original signal in time while keeping the overall duration intact. The performance curve (explored at multiple compression ratios and window sizes but shown to match human experiments) is estimated at 10 audio-to-silence duration ratios ranging from 0.5 to 2.0 on a logarithmic scale.

\subsection{Neural network models}
We evaluate a set of state-of-the-art speech recognition systems with diverse architectures and input types (Table \ref{table:1}; available through the cited references below).
These include fully-trained convolutional, recurrent, and transformer-based, with front ends that interface with either waveform or spectrogram inputs. 
Their accuracy under natural (unperturbed) conditions is high ($\sim 80\%$ correct, under word error rate)  and comparable (see e.g., Fig. \ref{figure:convergence}E when the warp factor equals 1, i.e., no perturbation).

\begin{table}[!ht]
\begin{center} 
\caption{Algorithmic models.} 
\label{table:1} 
\vskip 0.12in
\begin{tabular}{lll} 
\hline
Model    &  Architecture    &   Input\\
\hline
deepspeech  &   LSTM &  Spect.\\
wav2vec 2.0 &   1DConv-TF. &   Wave\\
fairseq-s2t &   2DConv-TF. &   Spect.\\
silero  &   Sep-2DConv.    &   Spect.\\
\hline
\end{tabular} 
\end{center}
\end{table}

\textit{Deepspeech} is based on a recurrent neural network architecture that works on the MFCC features of a normalized spectrogram representation \cite{hannun2014DeepSpeechScalingac}. This type of Long-short-term-memory (LSTM) architecture emerged as a solution to the problem of modeling large termporal scale dependencies. 
The input is transformed by convolutional, recurrent and finally linear layers projecting into classes representing a vocabulary of English characters. 
It was trained on the \textit{Librispeech} corpus \cite{panayotov2015LibrispeechASRCorpus2iicasspi} using a CTC loss \cite{gravesConnectionistTemporalClassification2006}.

\textit{Silero} works on a short-time Fourier transform of the waveform, obtaining a tailored spectrogram-like representation that is further transformed using a cascade of separable convolutions \cite{veysov2020towardimagenetstt}.  
It was trained using a CTC loss \cite{gravesConnectionistTemporalClassification2006} on the \textit{Librispeech} corpus, with alphabet letters as modeling units.

The \textit{Fairseq-S2T} model is transformer-based and its front end interfaces with log-mel filterbank features \cite{wang2020FairseqS2TFastace}.
It is an encoder-decoder model with 2 convolutional layers followed by a transformer architecture of 12 multi-level encoder blocks.
The input is a log-mel spectrogram of 80 mel-spaced frequency bins normalized by de-meaning and division by the standard deviation.
This architecture is trained on the \textit{Librispeech} corpus using a cross-entropy loss and a unigram vocabulary.

\textit{Wav2vec2} is a convolutional- and transformer-based architecture \cite{baevskiWav2vecFrameworkSelfSupervised, schneider2019Wav2vecUnsupervisedPretrainingac}.
As opposed to the previous architectures, it works directly on the waveform representation of the signal and it was pretrained using self-supervision.
In this case, the relevant features are extracted by the convolutional backbone, which performs convolution over the time dimension. 
The temporal relationships are subsequently modeled using the transformer's attention mechanism.
The input to the model is a sound waveform of unit variance and zero mean. It is trained via a contrastive loss where the input is masked in latent space and the model needs to distinguish it from distractors. To encourage the model to use samples equally often, a diversity loss is used in addition. The fine tuning for speech recognition is done by minimizing a CTC loss \cite{gravesConnectionistTemporalClassification2006} with a vocabulary of 32 classes of English characters. The model was trained on the \textit{Librispeech} corpus.

\subsection{Evaluation}
We measure the number of substitutions $S$, deletions $D$, insertions $I$, and correct words $C$, and use them to compute the word error rate (WER) reflecting the overall performance of models on speech recognition, as follows:

\begin{equation}
\begin{aligned}
    WER = \frac{S + D + I}{S + D + C}
\end{aligned}
\end{equation}

A lower score indicates fewer errors overall and therefore better performance. Since $N_{ref} = S + D + C$ is the number of words in the ground truth labels and it appears in the denominator, the WER can reach values greater than 1.0.

We evaluate all models on the \textit{Librispeech} test set \cite{panayotov2015LibrispeechASRCorpus2iicasspi}, which none of the models have seen during training, manipulated and resynthesized for each experiment according to our synthesis procedures. In all cases we plot average performance scores across this large set of utterances; with negligible variability. We use English language speech, as the effects we focus on in humans appear to be independent of language (e.g., \cite{gotoh2017EffectPermutationsTimetjotasoa}).

\subsection{Input statistics}
\subsubsection{Sparsity.} The input samples in the natural and synthetic versions of the evaluation set are characterized by their sparsity in time and frequency. We compute the Gini coefficient $G$ on an encoding of signal a $\va{x}$ of length $n$ (time or frequency representation), which exhibits a number of desirable properties as a measure of signal sparsity \cite{hurley2009ComparingMeasuresSparsityacm}.
\begin{equation}
\begin{aligned}
G = \sum_{i}^{n} \sum_{j}^{n} \frac{|x_i - x_j|}{2 n^2 \bar{x}}
\end{aligned}
\end{equation}

\noindent We characterize the joint distributions of sparsity in the time and frequency domain from the point of view of audition systems, which process sounds sequentially over restricted timescales. 
Specifically, we compute a time-windowed Gini, $G_w$, at various window lengths $w$, resulting in a multiple-timescale dynamic sparsity measure. We focus on the 220 ms timescale which roughly aligns with both human cognitive science results \cite{poeppel2003AnalysisSpeechDifferentsc} and receptive field sizes of neural network models. 
The result for a given timescale is summarized by statistics on the Gini coefficients across $n$ signal slices of length $w$:

\begin{equation}
\begin{aligned}
\hat{G_w} = f(\{G(\vb{x}^i)\}_i^n)
\end{aligned}
\end{equation}

\noindent where $f(.)$ may be the mean (our case), standard deviation, etc.
We obtain in this way both a time sparsity and a frequency sparsity measure for each speech utterance, for all natural and perturbed test signals.
Since $G$ is sensitive to the experimental manipulations, this allows us to summarize and visualize a low-dimensional, interpretable description of the distributions at the input of the systems (e.g., \citeNP{evans2021BiologicalConvolutionsImproveb}).

\section{Acknowledgments}

We thank Oded Ghitza for clarifications on the original repackaging experiments and Franck Ramus for providing information on various replications and extensions. 
We thank 3 anonymous reviewers for constructive feedback that allowed us to improve a previous version of the manuscript. 
This project has received funding from the European Research Council (ERC) under the European498 Union’s Horizon 2020 research and innovation programme (grant agreement No 741134), and the Ernst Strüngmann Foundation.

\bibliographystyle{apacite}

\setlength{\bibleftmargin}{.125in}
\setlength{\bibindent}{-\bibleftmargin}

\bibliography{ann_timescales}

\end{document}